%% file: paper.tex
\documentclass[conference]{IEEEtran}
\pdfoutput=1
\ifCLASSINFOpdf
  \usepackage[pdftex]{graphicx}
\else
\fi
\usepackage[cmex10]{amsmath}
\usepackage{url}
\usepackage[numbers]{natbib}

\usepackage{anyfontsize}
\usepackage{flushend}

\usepackage{color}

\begin{document}
%
\title{Building a RAPPOR with the Unknown:\\
\mbox{\fontsize{18}{20}\selectfont Privacy-Preserving Learning of Associations and Data Dictionaries}}

\author{\IEEEauthorblockN{Giulia Fanti}
\IEEEauthorblockA{
University of California-Berkeley \\
\texttt{gfanti@eecs.berkeley.edu}}
\and
\IEEEauthorblockN{Vasyl Pihur}
\IEEEauthorblockA{Google, Inc.\\
\texttt{vpihur@google.com}}
\and
\IEEEauthorblockN{{\'{U}l}far Erlingsson }
\IEEEauthorblockA{Google, Inc.\\
\texttt{ulfar@google.com}}
}

%


\maketitle

\begin{abstract}
Techniques based on randomized response enable the collection of potentially sensitive data from clients
in a privacy-preserving manner with strong local differential privacy guarantees. One of the latest such technologies, RAPPOR~\cite{RAPPOR},
allows the marginal frequencies
of an arbitrary set of strings to be
estimated via privacy-preserving crowdsourcing.
However, this original estimation
process requires a known set of possible strings;
in practice, this dictionary can often
be extremely large and sometimes completely unknown.

In this paper, we propose a novel decoding algorithm for the RAPPOR mechanism that enables the estimation of ``unknown unknowns,'' i.e., strings we do not even
know we should be estimating.
To enable learning without explicit knowledge of the dictionary, we develop methodology
for estimating the joint distribution of two or more variables collected with RAPPOR.
This is a critical step towards understanding relationships between
multiple variables collected in a privacy-preserving manner.

\end{abstract}


%
\IEEEpeerreviewmaketitle

\section{Introduction}
\input{intro.tex}

\begin{figure*}[htbp]
\begin{center}
\includegraphics[width = 6.5in]{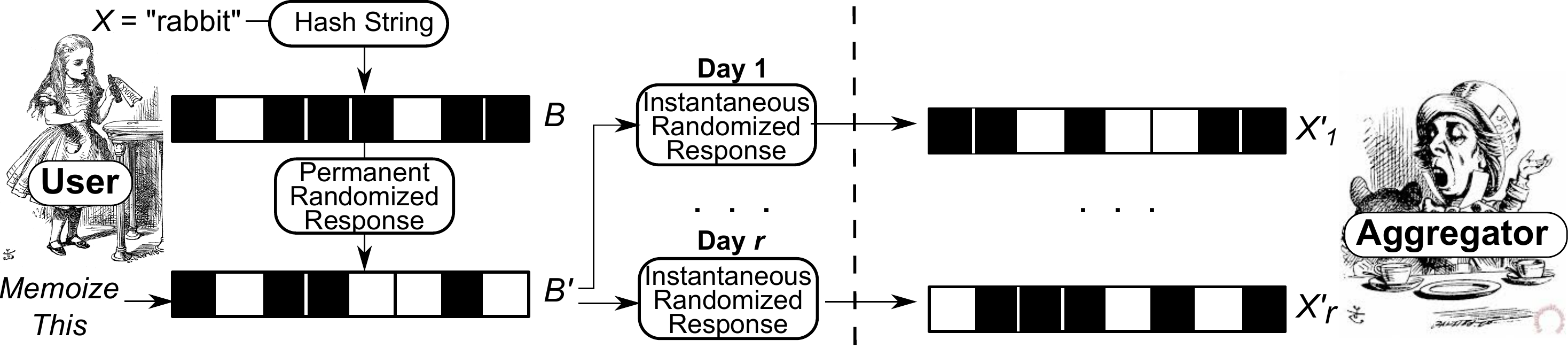}
\caption{Visualization of the RAPPOR mechanisms from \cite{RAPPOR}.
Each user starts by hashing her true string into a Bloom filter $B$.
This representation $B$ is used to generate a permanent randomized response (PRR) $B'$.
Each time the aggregator collects data (this might happen daily, for instance), the user builds a new instantaneous randomized response $X'$ from $B'$ and sends it to the aggregator.}
\label{fig:rappor}
\end{center}
\end{figure*}

\input{background.tex}

\input{regression.tex}

\section{RAPPOR Without a Known Dictionary} \label{sec:dictionary}
Suppose we wish to use RAPPOR to learn the ten most visited URLs last week.
To do this, we could first create an exhaustive list of candidate URLs,
and then test each candidate against received reports to determine which ones are present in the sampled population.
In this process, it is critical to include \emph{all} potential candidates, since RAPPOR has no direct feedback mechanism
for learning about missed candidates.
Such a candidate list may or may not be available, depending on what is being collected.
For instance, it may be easy to guess the most visited URLs, but if we instead wish to learn the most common tags in private photo albums, 
it would be impractical to construct a fully exhaustive list.
In this section, we describe how to learn distribution-level information about a population without knowing the dictionary, i.e., the set of candidate strings, beforehand.

To enable the measurement of unknown strings, more information needs to be collected from clients.
In addition to collecting a regular RAPPOR report of the client's full string, we will collect RAPPOR reports generated from $n$-grams\footnote{An
$n$-gram is an $n$-character substring.} that are randomly selected from the string.
The key idea is to use co-occurrences among $n$-grams to construct a set of full-length candidate strings. To analyze these co-occurrences, we use the joint distribution estimation algorithm developed in the previous section. Once we build a dictionary of candidate strings, we can perform regular, marginal RAPPOR analysis on the full-string reports to estimate the distribution.
In the extreme case, if our $n$-grams were as long as the string itself, we would be searching for candidates over the space of every possible string of a given length.
By using small $n$-grams (2 or 3 characters long), we can significantly reduce the associated computational load, without compromising accuracy.

Concretely, a client reporting string $x$ with local differential privacy budget $\epsilon$ would create a report
$$
X' = \text{RAPPOR}(x)
$$
by spending a third of her privacy budget (i.e., using differential privacy level $\epsilon / 3$). The other two thirds of $\epsilon$ would be spent equally on collecting
two $n$-grams 
$$
G_1' = \text{RAPPOR}(\text{$n$-gram}(x, g_1))
$$
and 
$$
G_2' = \text{RAPPOR}(\text{$n$-gram}(x, g_2))
$$
at distinct random positions $g_1$ and $g_2$, where $n$-gram$(x,g_i)$ denotes the length-$n$ string starting at the $g_i$th character. In principle, the only limitations on $g_1$ and $g_2$ are that $g_1\neq g_2$ and $g_1,g_2 \leq M - n$; this means that one could choose partially overlapping $n$-grams. In our simulations, we impose the condition that $M$ is divisible by both $g_1$ and $g_2$, meaning that we partition the string into adjacent, non-overlapping $n$-grams. For instance, if our strings have at most $M=6$ characters and our $n$-grams are two characters each, then there are only 3 $n$-grams per string; $g_1$ and $g_2$ are therefore drawn without replacement from the set $\{1,2,3\}$. 
In the original RAPPOR paper, each client would report a single randomized bit array $X'$.
Our proposed augmented collection would instead report $\{X', G_1', G_2', g_1, g_2\}$, where both $g_1$ and $g_2$, the two $n$-gram positions,
are sent in the clear.

To prevent leakage of information through the length of the string, $x$, the aggregator should
specify a maximum string length $M$
(divisible by the size of $n$-grams) and pad
all strings shorter than $M$ with empty spaces.
Strings longer than $M$ characters would be truncated and hashed to
create $X'$, and only $M / n$ distinct $n$-grams would be sampled.
Information in the tail of strings longer than $M$ would be permanently lost and can only be recovered
by increasing $M$.
There are interesting trade-offs involved in the selection of $M$, which should become clear after we describe the decoding algorithm.

Note that there is nothing wrong with using overlapping $n$-grams; for a fixed number of sampled $n$-grams, it increases redundancy at the expense of coverage, much like the use of overlapping windows in spectral signal analysis. 
Similarly, there is nothing theoretically wrong with measuring more than two $n$-grams. However, this would force each $n$-gram to use privacy level $\epsilon/(r+1)$, where $r$ is the number of $n$-grams measured; this forces the client to send more data to achieve the same fidelity on a per-$n$-gram basis.
More problematically, using larger numbers of $n$-grams can significantly increase the complexity of estimating $n$-gram co-occurrences.
We will discuss these details momentarily, but just to give an example, collecting 3 bigrams over the space of only letters requires us to estimate a distribution over a sample space with $(26^2)^3$ possibilities.
For this reason, we do not provide simulation results based on collecting more than two $n$-grams.

\begin{figure*}[htbp]
\begin{center}
\includegraphics[width = 6in]{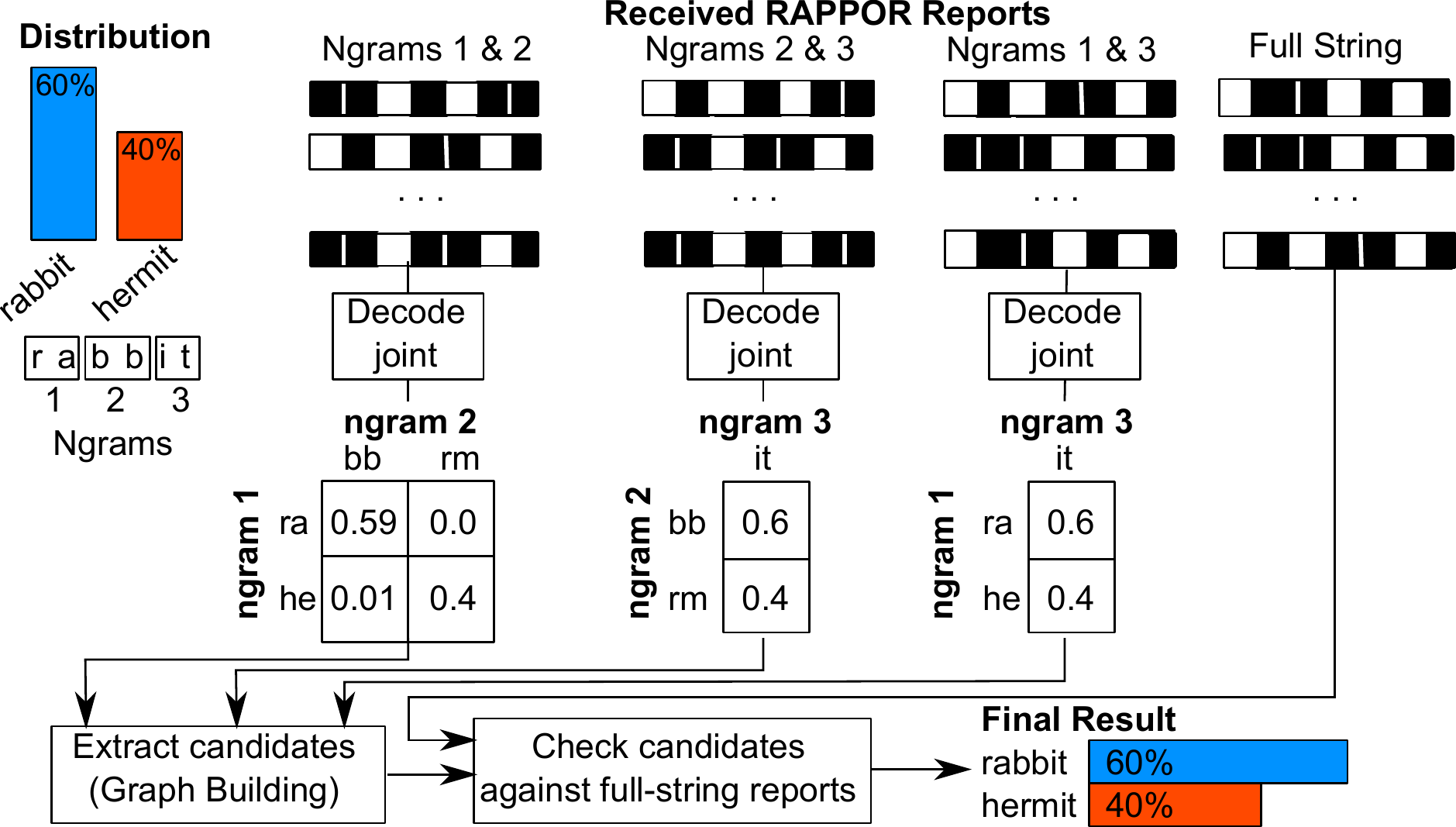}
\caption{Process for learning the distribution of a random variable without knowing the
dictionary ahead of time. The aggregator computes pairwise joint distributions from the
noisy reports generated at different $n$-gram positions.
These pairwise joint distributions are used to generate a candidate string dictionary.}
\label{fig:unknowns}
\end{center}
\end{figure*}

\subsection{Building the Candidate Set}

Let $N$ be the number of clients participating in the collection. The aggregator's reconstruction algorithm proceeds as follows:
\begin{enumerate}
\item {\bf Build $n$-gram dictionary: } Start by building a subdictionary of every possible $n$-gram. If the alphabet has $D$ elements in it, this subdictionary will have $D^n$ elements. An example alphabet is $D=\{0-9, a-z, -, \_, . \}$.
\item {\bf Marginal preprocessing: }Take the set of all reports generated from $n$-grams, $\{(G_1')_i,~(G_2')_i\}_{i=1}^N$. Split this set into mutually exclusive groups based on the position from which they were sampled.
There will be  $M / n$ such groups.
\item {\bf Marginal decoding:} For each position group, perform marginal analysis to estimate which $n$-grams are common at each position and their corresponding
frequencies. This step uses the $n$-gram dictionary constructed in (1).
\item {\bf Joint preprocessing:} Each pair of $n$-grams falls into one of ${M/n \choose 2}$ groups, defined by the randomly-chosen positions of the two $n$-grams, $g_1$ and $g_2$. Split the reports into these groups.
\item {\bf Joint analysis:} Perform separate joint distribution analyses for each group in (4) using the significant $n$-grams discovered in (3).
\item {\bf $n$-gram candidates:} Select all $n$-gram pairs with frequency greater than some threshold $\delta$.
\item {\bf String candidates (Graph-building):} Construct a graph with
edges specified by the previously-selected $n$-gram pairs. Analyze the graph to select all $M/n$-node fully connected subgraphs which form a candidate set $C$.
\end{enumerate}

Steps (3)--(7) are illustrated in Figure \ref{fig:unknowns}, but steps (6) and (7) require some more explanation.
For simplicity assume that $M = 6$ and that we are collecting
two bigrams from each client.
For string $x$ with frequency $f(x)$, there could only be three different combinations of bigram pairs reported by each client: $(g_1,g_2)\in\{(1, 2),~(1, 3),~(2, 3)\}$.
If string $x$ is a true candidate, then we would expect the corresponding bigrams from \emph{all three} pairings to have frequency of at least $f(x)$ in the relevant joint distributions.
Additional frequency could come
from other strings in the dictionary that share the same bigrams.
In general, \emph{all} $n$-gram pairs must have
frequency greater than some threshold $\delta$ to produce a valid candidate.
We computed $\delta$ as
\begin{equation*}
\delta = \sqrt{\frac{p_2 (1 - p_2)}{(q_2 - p_2)N}},
\label{eq:delta}
\end{equation*}
where
$$
q_2 = 0.5 f (p + q) + (1 - f) q
$$
and
$$
p_2 = 0.5 f (p + q) + (1 - f) p.
$$
This expression is designed to ensure that if an $n$-gram pair has no statistical correlation, then with high probability its estimated probability will fall below $\delta$. Indeed, $1.64 \delta$ is a frequency threshold above which we expect to
be able to distinguish strings from noise in our marginal analysis.
We deliberately use a slightly lower threshold to reduce our false negative rate.

\begin{figure}[htbp]
\begin{center}
\includegraphics[width = 2.7in]{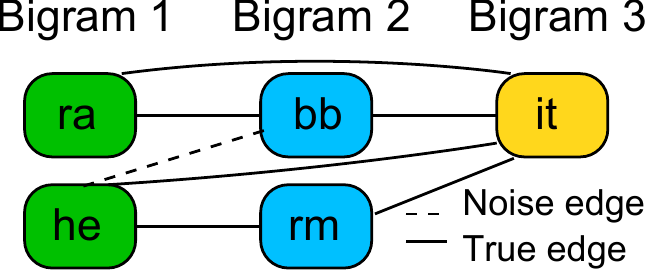}
\caption{Graph building process for generating full string candidates.
In the graph-building phase, we search for fully connected cliques in this $k$-partite graph.
In this graph, the resulting set of candidate strings would be $C=\{\text{rabbit},~\text{hermit},~\text{hebbit}\}$.
The noisy false positive (``hebbit") gets weeded out by candidate testing (section \ref{sec:test}).}
\label{fig:graph}
\end{center}
\end{figure}

Step (7) is explained in greater detail in Figure \ref{fig:graph}.
The basic idea is to construct a set of candidate strings by building a graph and finding fully-connected cliques.
Each $n$-gram at each position is treated as a distinct node.
Edges are drawn between every valid $n$-gram pair from step (6).
These edges may be due to true signal (solid lines) or noise (dotted lines), but the aggregator has no way of distinguishing \emph{a priori}.
Regardless of provenance, edges are only drawn between $n$-grams of different positions, so the resulting graph is $k$-partite, where $k=M/n$.
Now the task simplifies to finding every fully-connected $k$-clique in this $k$-partite graph; each clique corresponds to a candidate string.
This works for the following reason: If a string $x$ is truly represented in the underlying distribution, then the likelihood of \emph{any} $n$-gram pair
having a joint distribution below the threshold $\delta$ is small.
Therefore, if even a single $n$-gram pair from string $x$ has a significantly lower frequency than $\delta$
after accounting for the noise introduced by RAPPOR, then it is most likely a false positive.
Accordingly, the corresponding edge will be missing in the graph, and
our clique-finding approach will discard $x$ as a candidate string.

If executed naively, this clique-finding step can become a storage and computation bottleneck.
In the worst case, the number of candidates can grow exponentially in the number of bigrams collected.
The problem of efficiently finding $k$-cliques in a $k$-partite graph has been studied in the context of braiding
in the textile industry \cite{mirghorbani2013finding}; this approach significantly outperforms traditional branch-and-bound algorithms.

Candidates in $C$ can be further filtered based on string semantics and/or limited prior knowledge.
For example, if it becomes apparent that what we are collecting are URLs, then candidates that do not meet strict URL encoding restrictions can
be safely removed without further consideration (e.g., strings with spaces in the middle and so on).

\subsection{Testing Candidate Strings}\label{sec:test}
To estimate the marginal distribution of unknown strings, we use the set of full string reports $X'_1, \ldots, X'_N$ and candidate dictionary $C$ 
to perform marginal inference as described in the original RAPPOR paper.
False positives in the candidate set $C$ will be weeded out in this step, because the marginal decoding
shows that these strings occur with negligible frequency.
The marginal analysis here differs from classical RAPPOR marginal analysis in two important ways:
\begin{enumerate}
\item Reports $X'_1, \ldots, X'_N$ are collected with stronger privacy guarantees by using privacy parameter $\epsilon / 3$ as opposed to $\epsilon$.
Depending on the true distribution, this may or may not affect the final results, but in general, there is a substantial penalty for collecting
additional information in the form of two $n$-grams.
\item The estimated candidate set $C$ is unlikely to be as complete as an external knowledge-based set.
With high probability, it will include the most frequent (important) candidates, but it will miss less frequent strings due to privacy guarantees imposed on $n$-gram reporting.
On long-tailed distributions, this means that a significant portion of distribution mass may fall below the noise floor.
Set $C$ is also likely to be comprised of many false-positive candidates forcing a higher stress load on statistical testing that
necessarily must be more conservative in the presence of a large number of tests.
\end{enumerate}
The output of this step is the estimated marginal weights of the most common strings in the dictionary.

\section{Results}
We performed a series of simulation studies and one real-world example to empirically show the utility of the proposed approach. Before showing these results, we discuss why this scheme does not alter the privacy guarantees of original RAPPOR. 

\subsection{Privacy}
Recall that we split the privacy budget evenly between the $n$-grams and the full-string report. For instance, if we collect reports on two $n$-grams and the full string, each report will have privacy parameter $\epsilon/3$.
It is straightforward to show from the definition of local differential privacy that two independent measurements of the same datapoint, each with differential privacy parameter $\gamma$, will collectively have privacy parameter $2\gamma$.
Moreover, dependent measurements contain less information than independent measurements, so the overall privacy parameter is \emph{at most} $2\gamma$.
Consequently, our $n$-gram based measurement scheme provides the same privacy as a single RAPPOR report with differential privacy $\epsilon$. 

Note also that local differential privacy guarantees hold even when the aggregator has side information \cite{kasiviswanathan2008note}. For instance, the aggregator might wish to study a distribution of strings from a small (but unknown) dictionary of English words; it might therefore have a prior distribution on bigrams that appear in the English language. In this case, differential privacy guarantees ensure that the aggregator cannot improve its estimate (conditioned on the prior information) by more than a factor of $e^\epsilon$. 

\subsection{Efficiency}
Recall that $|D|$ is the size of our alphabet, and $r$ is the number of $n$-grams collected from each string. $N$ denotes the number of datapoints.
The bottleneck of our algorithm is constructing the dictionary of candidate strings.
This can be split into two phases: (a) computing $n$-gram co-occurrences, and (b) building the candidate dictionary from a $k$-partite graph of $n$-gram co-occurrences.
Part (a) has complexity $O(N|D|^{nr})$ due to the EM algorithm.
Part (b) depends on the size of the initial $k$-partite graph. If there are $p$ nodes in each of the partitions, this part has worst-case computational complexity $O(kp^{k-1})$.
However, due to the significant sparsity in this $k$-partite graph, this complexity can be much lower in practice.

These asymptotic costs can be prohibitive as the number of data samples increases.
This is partially because the EM algorithm in phase (a) is iterative, and each iteration depends on every data element; this can lead to high memory constraints and lengthy runtimes.
However, while the complexity of part (a) dominates part (b) in most usage scenarios, part (a) can also be parallelized more easily. 
We are therefore working on releasing a parallelized version of the EM estimation code. 
We will also show how the parameter $\delta$ can be tuned to reduce the computational load of part (b) in exchange for a reduction in accuracy.

\subsection{Simulated Results}
To understand the impact of parameter choices on accuracy and efficiency, we built a synthetic dataset comprised of
fake ``hashes''---randomly selected character strings of a fixed length.
We then specified a distribution over 100 such strings;
in the following examples, that distribution is a discretized Zipfian.\footnote{Data
generated from other distributions are included in Appendix \ref{app:distributions}.}
We drew 100,000 strings from this distribution, and encoded them as 128-bit RAPPOR reports, with parameters 

\begin{figure}[htbp]
\begin{center}
\includegraphics[width = 3.7in]{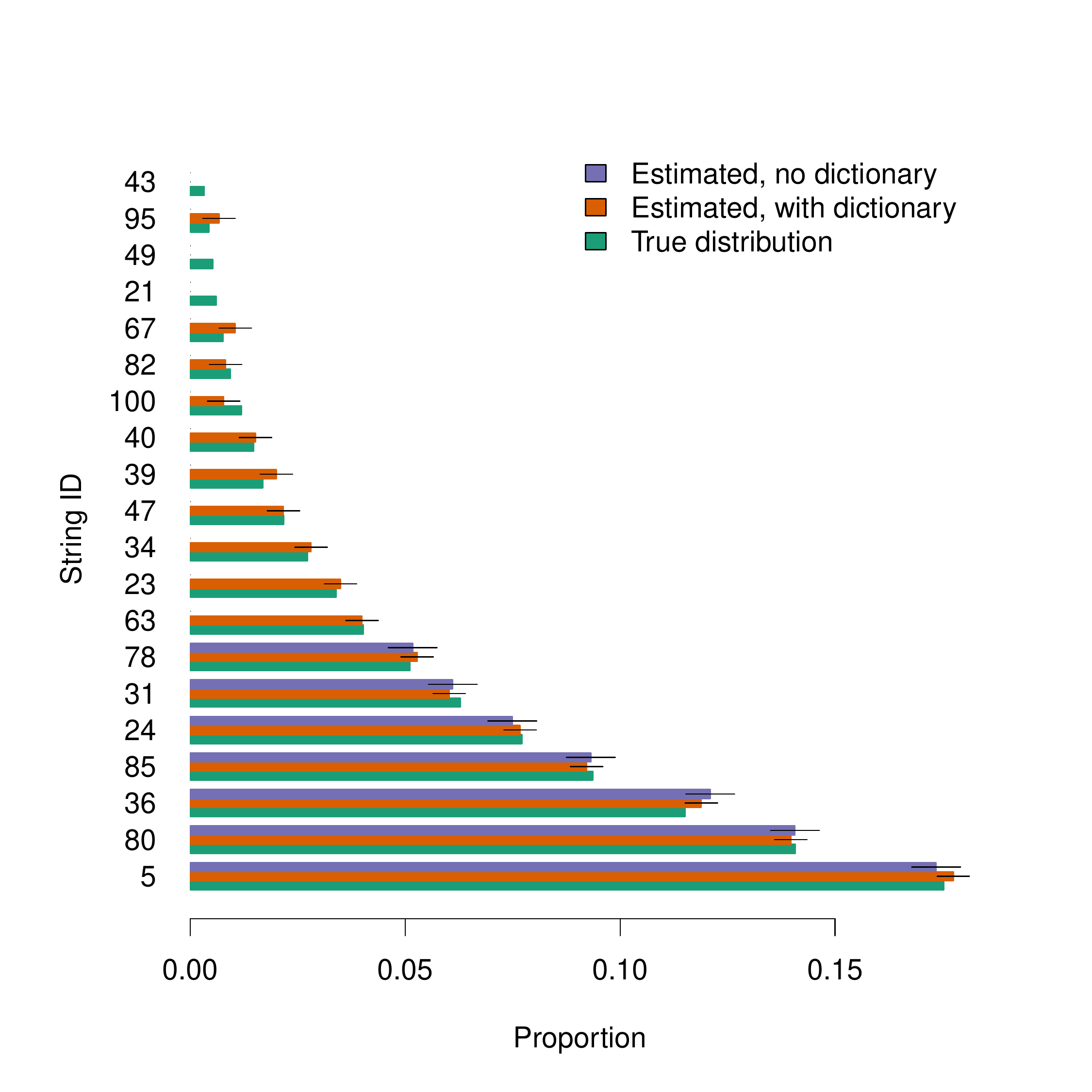}
\caption{Estimated distribution of hash strings,
computed from the RAPPOR reports of 100,000 simulated users.
This plot only shows results for the 25 most frequent strings.
We were able to correctly estimate the top 7 strings without access to a prior dictionary.
With the dictionary as prior information, we are able to estimate 17 of the top strings.}
\label{fig:mean}
\end{center}
\end{figure}

$p=0.25$, $q=0.75$, and $f=0$.
Based on this set of noisy reports and our joint analysis, we estimated the marginal distribution of these strings
without using \emph{any} prior knowledge about them.
We limited ourselves to 100,000 strings for the sake of computational feasibility while exploring the parameter space.
However, in this section we will also show results from a larger trial on a real dataset with 1,000,000 simulated clients.

Figure \ref{fig:mean} illustrates that our method correctly estimates the underlying distribution on average.

To evaluate how close our estimate was to the true distribution, we used the Hellinger distance, which captures the distance between
two distributions of discrete random variables.
For discrete probability distributions $P$ and $Q$ defined over some set $\mathcal U$,
the Hellinger distance is defined as
\[
H(P,Q) = \frac{1}{\sqrt{2}}\sqrt{\sum_{i\in \mathcal U}(\sqrt{P(i)}-\sqrt{Q(i)})^2}.
\]
This metric is related to the Euclidean norm of the distance between the square root vectors; we chose it in part because unlike
Kullback-Leibler divergence, it is defined
even when the two distributions are nonzero over different sets.

\begin{figure}[htbp]
\begin{center}
\includegraphics[width = 3.5in]{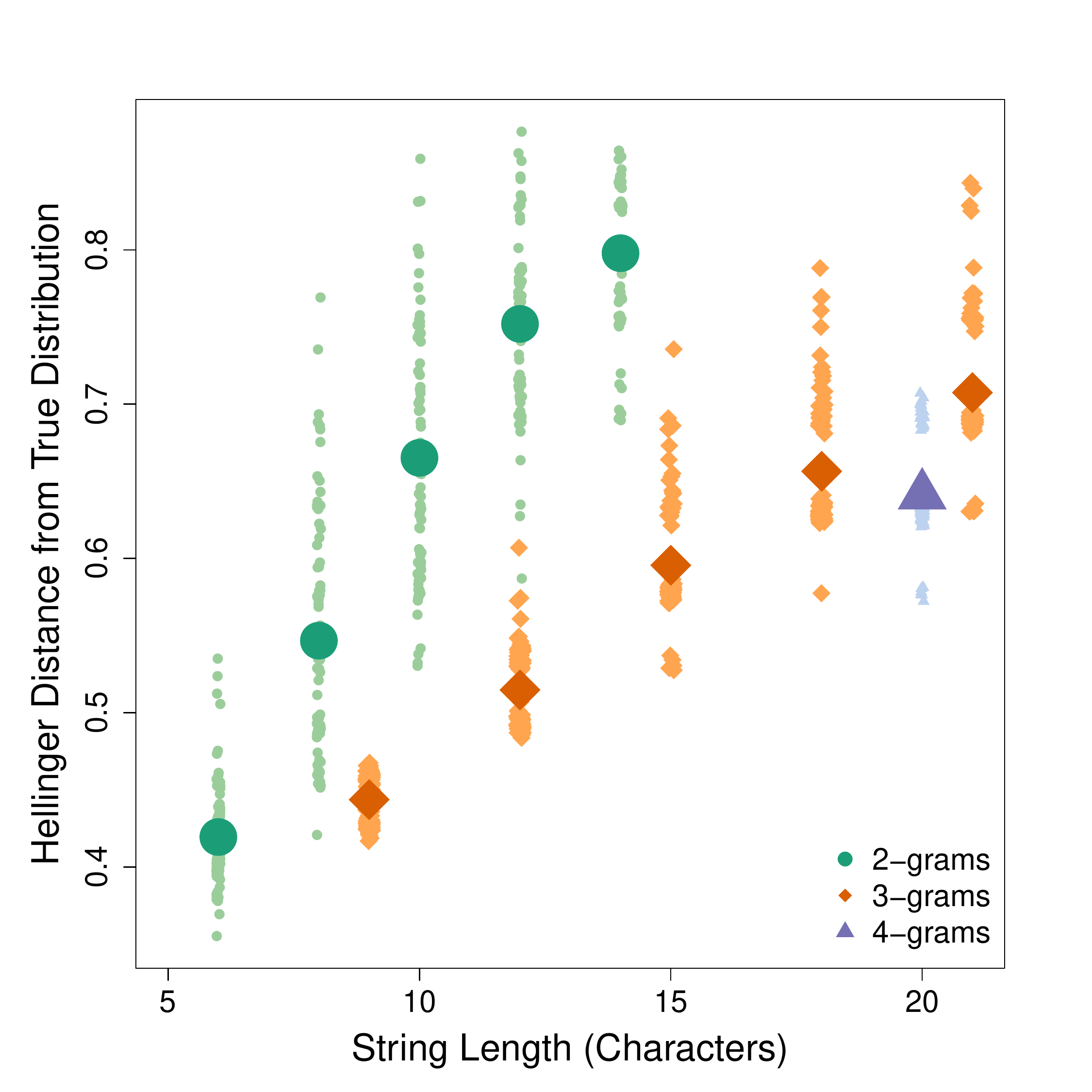}
\caption{Hellinger distance of our learned distribution from the true distribution as a function of string length.
The population size is 100,000 and each data point is averaged over 100 trials (illustrated in lighter face color).
Each user sent information on two $n$-grams from each string.}
\label{fig:divergence}
\end{center}
\end{figure}

\textbf{Accuracy and $n$-gram length: }
Figure \ref{fig:divergence} plots the Hellinger distance of our reconstructed distribution as a function of string length, for different
sizes of $n$-grams.\footnote{We only generated one point using 4-grams due to the prohibitive memory costs of decoding a dictionary with $26^4$ elements.}
This figure suggests that for a fixed string length, using larger $n$-grams gives a better estimate of the underlying
dictionary. Intuitively, this happens for two reasons: (1) Reports generated from longer $n$-grams contain information about a larger fraction
of the total string; we only collect two $n$-grams for communication efficiency, so the $n$-gram size determines what fraction of a string is
captured by reports. (2) The larger the $n$-gram, the fewer $n$-gram pairs exist
in a string of fixed length. In simulation, we observe that the likelihood of our algorithm missing an edge between $n$-grams is roughly constant,
regardless of $n$-gram size. Therefore, if there are more $n$-gram pairs to consider with smaller $n$-grams, the likelihood that at least one of the
edges is missing---thereby removing that string from consideration---is significantly higher.

\begin{figure}[htbp]
\begin{center}
\includegraphics[width = 3.5in]{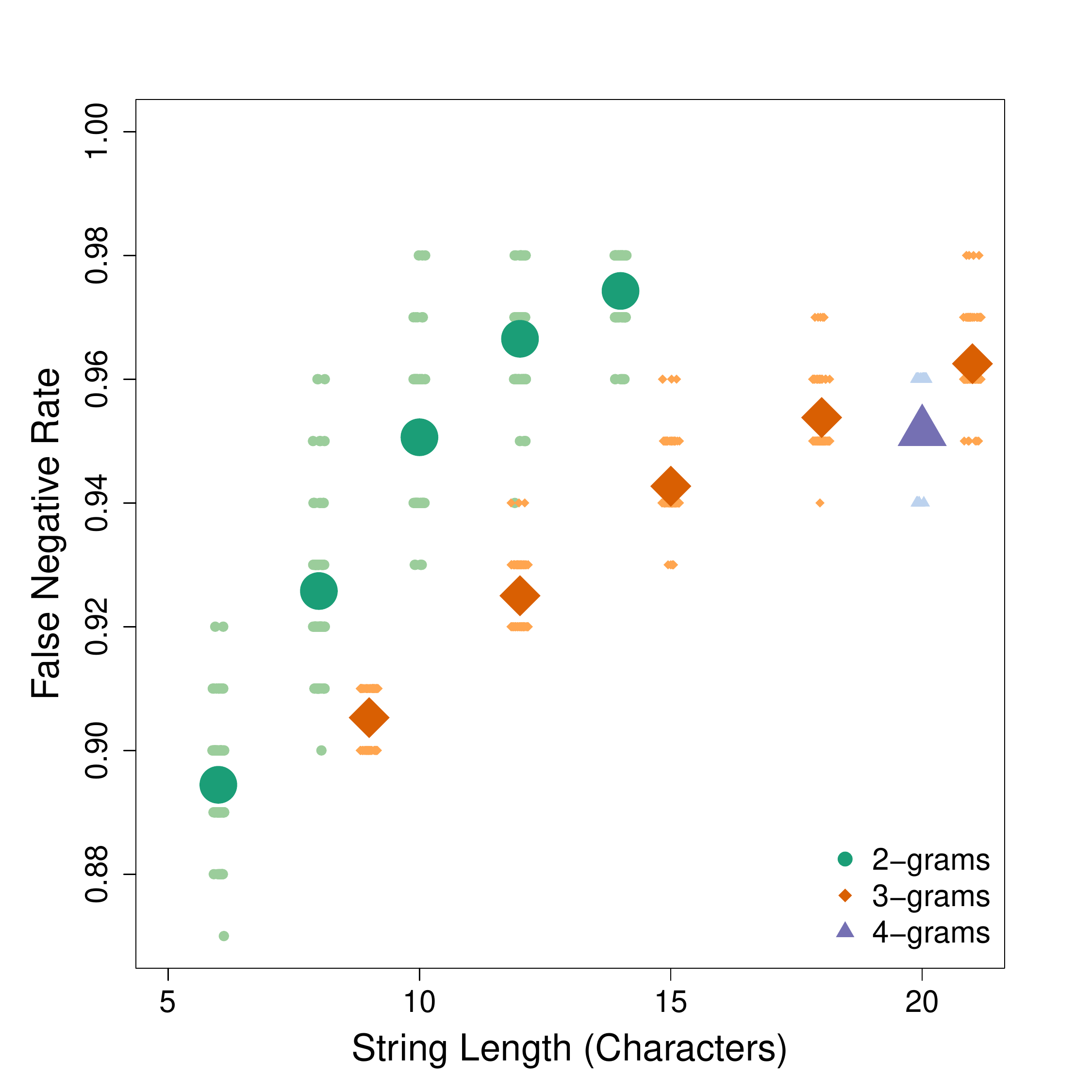}
\caption{False negative rate as a function of string length.
This was generated from the same data summarized in Figure \ref{fig:divergence}.
We observe that discrepancies between distributions are primarily caused by false negatives.}
\label{fig:fn}
\end{center}
\end{figure}

This hypothesis is supported by Figure \ref{fig:fn}, which shows the false negative rate as a function of string size for different $n$-gram sizes.
In all of these trials, we did not observe \emph{any} false positives, so false negatives accounted for the entire discrepancy in distributions.
Because our distribution was quite peaked (as is the case in many real-life distributions over strings), missing even a few strings caused the overall
distribution distance to decrease significantly.

\textbf{Accuracy vs. computational costs: }
As mentioned previously, graph-building becomes a bottleneck if the EM portion of the algorithm is properly parallelized and optimized.
This stems from the potentially large number of candidate strings that can emerge while searching over the $k$-partite graph of $n$-grams.
This number depends in part on the threshold $\delta$ used to select ``significant'' associations between $n$-grams.
Choosing a larger threshold results in fewer graph edges and lower computational load, but this comes at the expense of more missed strings in the
candidate set.

\begin{figure}[htbp]
\begin{center}
\includegraphics[width = 3.5in]{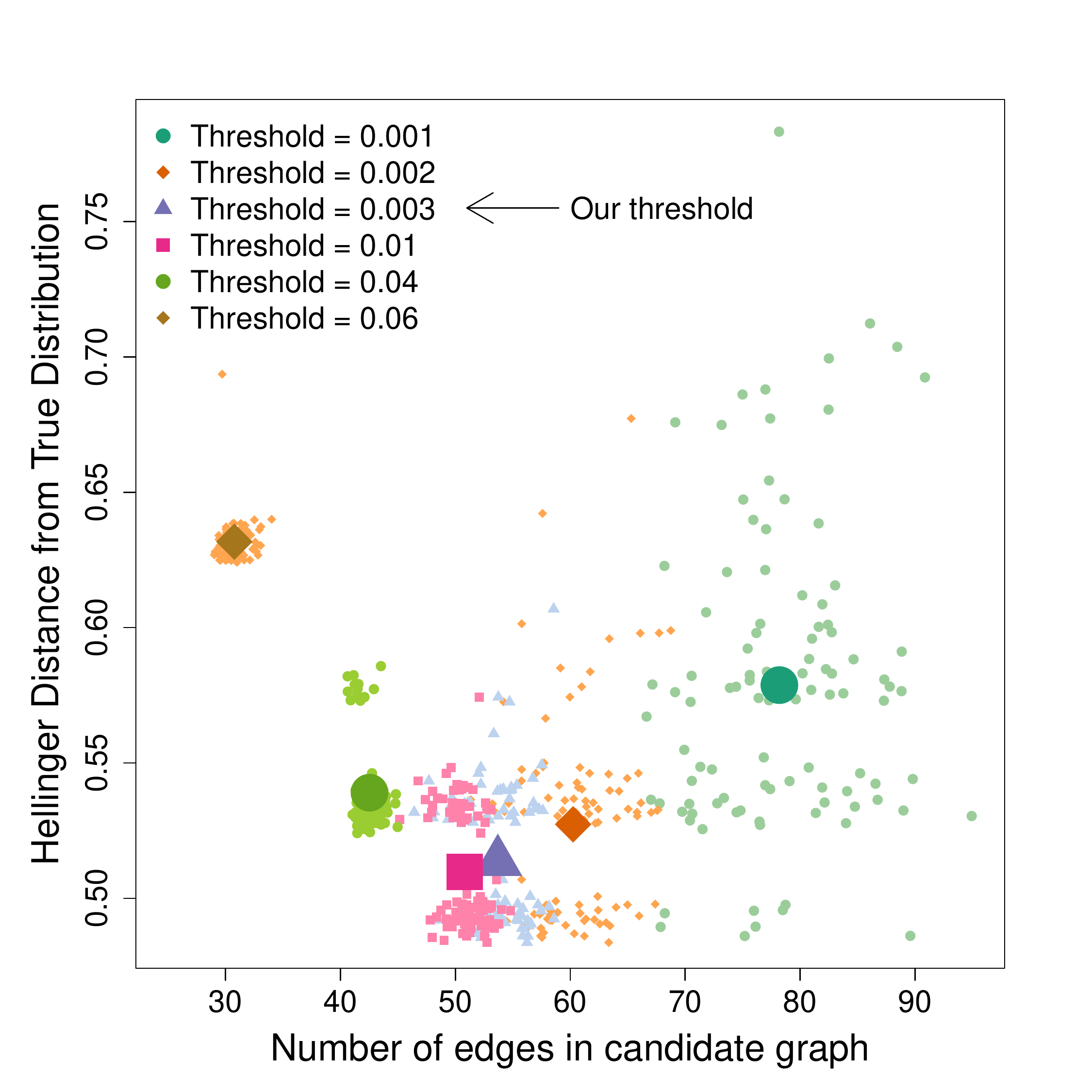}
\caption{The Hellinger distance between our recovered distribution and the true distribution as a function of the number of edges in the
k-partite candidate graph. Each cluster was generated using a different edge-creation thresholds in the pairwise joint distributions.
We observe that optimal threshold value is close to our chosen threshold, which
can be deterministically computed from the randomized response noise parameters.}
\label{fig:thresh}
\end{center}
\end{figure}

To understand this tradeoff better, we examined the impact of the pairwise candidate threshold on accuracy.
In principle, if we were to set the threshold to zero, we could recover every string in the dictionary.
However, this greatly increases the false positive rate, as well as the algorithmic complexity of finding those strings.
Figure \ref{fig:thresh} plots the Hellinger distance of the recovered distribution against the number of edges in the candidate $n$-gram graph for
various distribution thresholds.
The number of edges in the $n$-gram k-partite graph indirectly captures the computational complexity required to build and prune candidates.

As expected, the computational complexity (i.e. number of edges in the candidate graph) decreases as the threshold increases.
However, counterintuitively, the accuracy decreases for very low thresholds.
This occurs because each candidate string is treated as an independent hypothesis, the null hypothesis being that the candidate is not significant.
When testing for $M$ independent hypotheses with significance $\alpha$, it is common practice to use \emph{Bonferroni correction}, which
reduces the significance of each individual
test to $\alpha / M$ in order to account for the greater likelihood of seeing rare events when there are multiple hypotheses.
The net effect of this is to impose more stringent significance tests when there are more candidates.
Since lowering the threshold also increases the number of candidate strings, 
the resulting Bonferroni 
correction causes many true strings
to fail the significance test.
If we did not use Bonferroni correction, we would observe a high number of false positives.
Due to this effect, we observe a clear optimal threshold value in Figure \ref{fig:thresh}.
The optimal parameter setting is difficult to estimate without extensive simulations that depend on the distributional information
we're trying to estimate in the first place.
However, in simulation we observe that the threshold computed analytically in Eq. (\ref{eq:delta})---which is based on the statistics of
the randomized response noise---appears close to the optimum, and is likely a good choice in practice.

\subsection{Estimating the Dictionary in Real-World Settings}
%

To understand how this approach might work in a real setting, we located a set of 100 URLs
with an interesting real-world frequency distribution
(somewhat similar to the Alexa dataset~\cite{Alexa}).
We simulated measuring these strings through RAPPOR by drawing one million strings from the distribution, and
encoding each string accordingly.
We then decoded the reports using varying amounts of knowledge about the underlying dictionary of URLs.


All URL strings were padded with white space up to 20 characters, matching the longest URL in the set.
In addition to full string reports, two randomly-chosen bigrams (out of 10) were also reported, all using $128$-bit Bloom filters
with two hash functions.
Overall privacy parameters were set to $q = 0.75$, $p = 0.25$ and $f = 0$ (assuming one-time collection).
This choice of parameters provides $\epsilon = 4.39$ or $\exp(\epsilon) = 81$ privacy guarantees, deliberately set quite high for
demonstration purposes only.
Each of the collected reports---based on the string itself and two bigrams---were allotted equal privacy budgets of
$\epsilon / 3$, resulting in effective parameter choices of $p = 0.25$ and $q = 0.32$.

\begin{figure*}[htbp]
\begin{center}
\includegraphics[scale = .5]{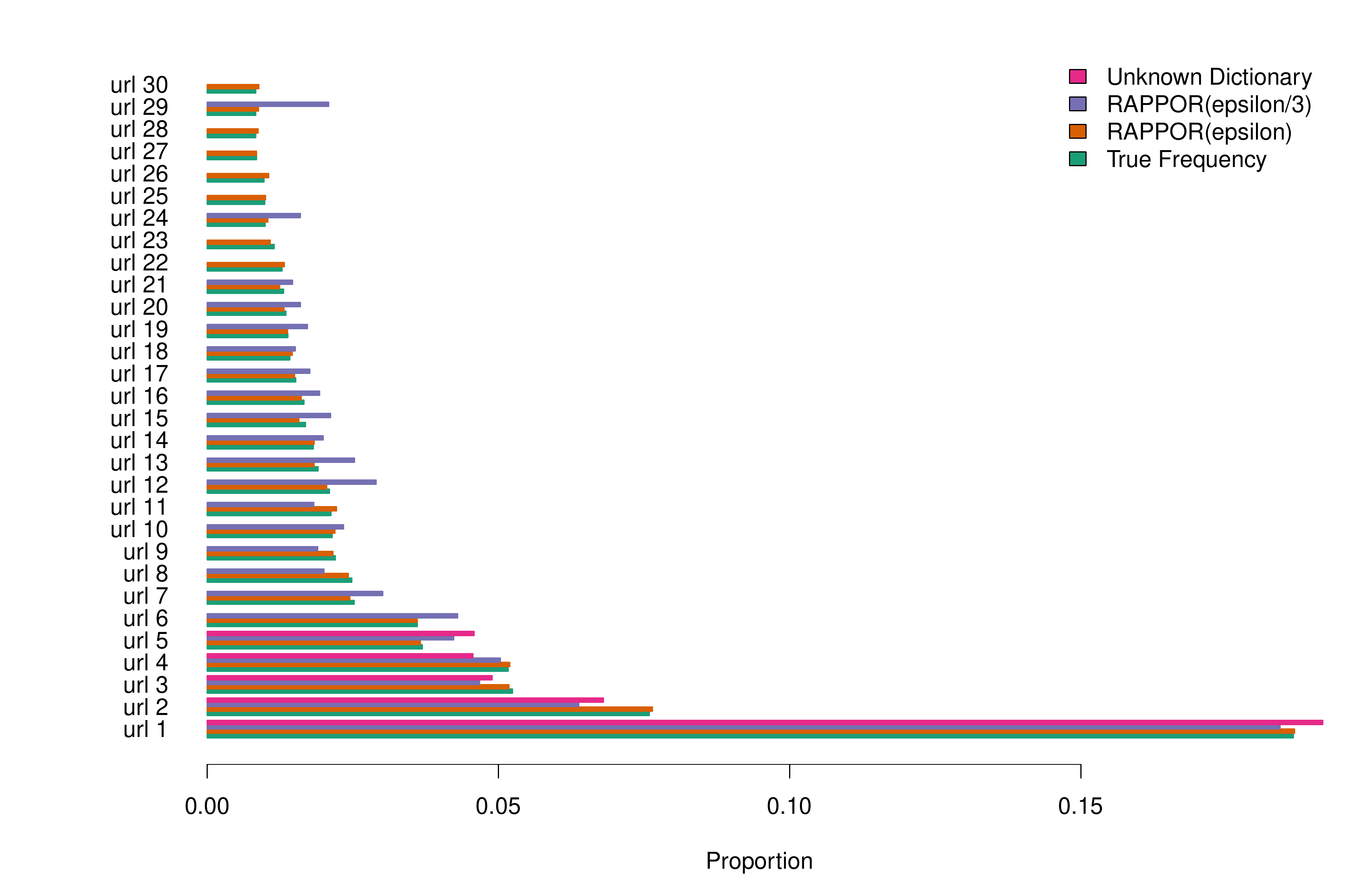}
\caption{Learning a distribution of URLs with and without the dictionary.
The dictionary contains 100 strings (only the top 30 are shown).
The bottom bar indicates the URL's true frequency.
The second bar from the bottom shows estimated frequencies when collected at $\epsilon = 4.39$ with full knowledge of the dictionary.
The third bar also assumes the full knowledge of the dictionary, but the collection took place with $\epsilon / 3 = 1.46$ (stronger privacy);
this illustrates losses incurred by allocating 2/3 of the privacy budget to collecting two bigrams.
The top bar shows distribution estimates computed without any knowledge of the dictionary;
each report and ngram was again encoded at $\epsilon / 3$ privacy.
}
\label{fig:search_engines}
\end{center}
\end{figure*}

Results are shown in Figure \ref{fig:search_engines}, where we truncate the distribution to the top 30 URLs for readability.
Each URL's true frequency is illustrated by the green bar.
The other three bars show frequency estimates for three different decoding scenarios.
A missing bar indicates
that the string was not discovered under that particular decoding scenario.

Under the first scenario, we performed an original RAPPOR analysis with $\epsilon = 4.39$ and perfect knowledge of the 100 strings in the dictionary.
With 1 million reports, we were able to detect and estimate frequencies for 75 unique strings.
The second scenario also assumes perfect knowledge of all 100 strings,
but performs collection at $\epsilon / 3 = 1.46$.
This illustrates how much we lose purely by splitting up the privacy budget to accommodate sending more information.
In this second scenario, 23 strings were detected, and their estimated frequencies are shown with blue bars.

In the third scenario, no prior knowledge of the dictionary was used.
Each string and bigram was collected at privacy level $\epsilon / 3 = 1.46$.
Ten marginal bigram analyses for each bigram position returned $4,~2,~2,~3,~4,~5,~2,~1,~1,$ and $1$ significant bigrams, respectively.
After conducting joint distribution analysis on pairs of bigrams, we selected bigram pairs whose joint frequency was above the threshold cutoff of $\delta = 0.0062$.
We then located the $10$-cliques in the corresponding $10$-partite graph, which produced 896 candidate strings.
The final marginal analysis based on the full string reports (to weed out false positives) discovered the top five strings and estimated their frequency quite accurately (pink bars).
There was also one false positive string identified by the analysis.
We also reran the collection with trigrams, which produced only 185 candidate strings.
Final marginal analysis resulted in only two strings with no false positives.

\textbf{A note on accuracy:}
Unfortunately, our method does not detect many of the strings in the population.
While we make no claims of optimality---either about the RAPPOR mechanism or our estimation algorithm---there is a well-studied fundamental tension between local differential privacy and data utility. 
Compared to estimating a distribution from $N$ unmasked samples, estimating it with locally-differential privacy reduces the effective sample size quadratically by $\epsilon$ when $\epsilon < 1$ \cite{duchi2013local}.
Since we collect each $n$-gram with privacy parameter $\epsilon/3$, our effective learning rate is slowed down significantly compared to regular RAPPOR.
Moreover, estimation over an unknown dictionary introduces an even greater challenge:
Worst-case, estimating a multinomial distribution at a given fidelity requires a number of samples that scales linearly in the support size of the distribution. 
So if we wish to estimate a distribution over an unknown dictionary of 6-letter words \emph{without} knowing the dictionary, in the worst case, we will need on the order of 300 million samples---a number that grows quickly in string length.
Considering these limitations, it is to be expected that learning over an unknown dictionary will perform significantly worse than learning over a known dictionary, regardless of algorithm.
Our algorithm nonetheless consistently finds the most \emph{frequent} strings, which account for a significant portion of the distribution's probability mass, both in our example and in many distributions observed in practice.
This enables an aggregator to learn about dominant trends in a population without any prior information and without violating the privacy of users.

\section{Related Work}
Since its introduction nearly a decade ago, 
differential privacy has become perhaps the best studied and most widely accepted definition of privacy~\cite{dwork2006differential}.
When there is no party
trusted to construct a database of the sensitive data,
the more refined notion of
local differential privacy 
is often considered~\cite{localdiffpriv,duchi2013local,kairouz2014extremal}.
Research on local differential privacy has largely been centered around finding algorithms that satisfy differential privacy properties \cite{mcsherry2009differentially,rastogi2010differentially,shi2011privacy},
and improving the tradeoffs between privacy and utility \cite{duchi2013local,kairouz2014extremal}.

Our work follows in a recent trend of using local differential privacy to learn a distribution's \emph{heavy-hitters}---the most significant categories in a distribution \cite{kilian2008fast,chan2012differentially,hsu2012distributed,RAPPOR}. Several of these papers focus on the information-theoretic limits of estimating heavy-hitters while satisfying differential privacy.
Our paper differs from existing work by asking new questions aimed at improving the practicality of the recently-introduced RAPPOR mechanisms~\cite{RAPPOR}.
Specifically, we consider two key questions: how to decode joint distributions from noisy reports, and how to learn distributions when the aggregator does not know the dictionary of strings beforehand.

Our work combats the recent notion that differential privacy is mainly of theoretical interest~\cite{bambauer2013fool}.
Therefore, we have identified two of the main technical shortcomings of a differentially-private mechanism 
that has seen practical, real-world deployment, namely RAPPOR, and
provided usable solutions that address those shortcomings.

The question of estimating distributions from differentially private data is not new,
with
Williams \emph{et al.} first making explicit the connection between probabilistic inference and 
differential privacy~\cite{williams2010probabilistic,jain2011differentially}.
This previous work is similar in principle to our approach.
There has even been some work on the distinct but related problem of releasing differentially private
marginal distributions generated from underlying multivariate
distributions~\cite{fienberg2011differential}.

However, existing work on distribution estimation from differentially private data 
only considers continuous random variables, 
while our work focuses on discrete random variables (specifically, strings).
This difference leads to significant practical challenges not addressed by prior literature,
and addressing those challenges is crucial to improving the RAPPOR mechanism.

Learning the distribution of random strings through differentially private data with an unknown dictionary is, to the best of our knowledge, a previously-unstudied question.

\section{Discussion}
Privacy-preserving crowdsourcing techniques
have great potential 
as a means of resolving the tensions
between individuals' natural privacy concerns
and the need to learn overall statistics---for
each individual's benefit, as well as for the common good.
The recently-introduced RAPPOR mechanism
provides early evidence that such techniques can be implemented in practice,
deployed in real-world systems,
and used to provide statistics with some benefits---at
least in the application domain of software security.
In this paper, we have addressed two significant limitations of this original RAPPOR system---namely
its inability to learn the associations between RAPPOR-reported variables,
and its need to known the data dictionary of reported strings ahead of time.
Notably, we have been able to achieve those
improvements without changing the fundamental RAPPOR mechanisms
or weakening its  local differential privacy guarantees.
%

This said, our new analysis techniques are not without their own shortcomings.
From a practical deployment perspective, the main limitation of our methods is the cost of decoding, 
where the primary bottleneck is the cost of joint estimation.
Parts of the joint distribution estimation algorithm are parallelizable, but each iteration of the EM algorithm ultimately
depends on the previous iteration, as well the entire dataset. If the number of users is very large, this can lead to significant memory and computational
loads.
The original RAPPOR system gets around this by using a LASSO-based decoding scheme that removes the dependency on every individual
report while returning unbiased estimates of the marginal distribution.
We cannot use this same approach because we do not have access to the joint count frequencies.
Nonetheless, a similar lightweight decoding algorithm for multivariate distributions would significantly improve the practicality of 
our enhanced RAPPOR analysis.

Another shortcoming in our current work
is the lack of optimal methods for parameter selection, and the allocation of privacy budgets.
In our experiments, we have allocated 1/3 of the privacy budget to each full-string report and two $n$-gram reports, but this allocation does not necessarily
maximize estimation accuracy.
For instance, it may provide more utility, at similar overall levels of privacy,
to collect $n$-grams with more relaxed privacy guarantees to get a
better estimate of the candidate set, and then use stricter privacy settings when collecting full
string reports.
Because our algorithm's performance is distribution-dependent, it is difficult to estimate optimal settings theoretically.
Moreover, searching over the complete parameter space is computationally challenging, due to the cost of decoding.
We hope that the eventual public release of our
analysis mechanisms (deferred for blind review)
will encourage experimentation on both fronts.
Furthermore, we also aim to tackle these challenges ourselves,
in future work.

\section{Conclusions}
Privacy-preserving crowdsourcing techniques
based on randomized response
can provide useful, new insights into unknown distributions
of sensitive data,
even while providing the strong guarantees
of local $\epsilon$-differential privacy.
As shown in this paper,
such privacy-preserving statistical learning
is possible even when there are multiple degrees and levels of unknowns.
In particular, 
by augmenting the analysis methods of the existing RAPPOR mechanism
it is possible to learn the joint distribution
and associations between two or more unknown variables,
and learn the data dictionaries of frequent, unknown values
from even very large domains, such as strings.
Furthermore,
those augmented RAPPOR analysis techniques
can be practical,
and can be of value when applied to real-world data.






%

\bibliographystyle{plainnat}
\bibliography{paper}

\input{appendices.tex}

\end{document}

%% file: intro.tex
It is becoming increasingly commonplace for companies and organizations to analyze user data in order to improve services or products.
For instance, a utilities company might collect water usage statistics from its users to help inform fair pricing schemes.
Although analyzing user data can be very beneficial, it can also negatively impact user privacy.  Data collection
can quickly form databases
that
reveal
sensitive details, such as preferences, habits, or personal characteristics
of implicitly or explicitly identifiable users.
It is therefore important to develop methods for analyzing the data of a population without sacrificing individuals' privacy.

A guarantee of \emph{local differential privacy} can provide the appropriate privacy protection
without requiring individuals to trust the intentions of a data aggregator~\cite{localdiffpriv}.
Informally, a locally differentially-private mechanism asks individuals to report data to which they have added carefully-designed noise,
such that any
individual's information cannot be learned, but an aggregator can correctly infer population statistics.
The recently-introduced
Randomized Aggregatable Privacy-Preserving Ordinal Response (RAPPOR) is the first such
mechanism 
to see real-world deployment~\cite{RAPPOR}.

RAPPOR is motivated by the problem of estimating a client-side distribution of string values drawn from a discrete data dictionary.
Such estimation is useful in many security-related scenarios.
For example,
RAPPOR is reportedly used in the Chrome Web browser
to track the distribution of
users' browser configuration strings; this is done to detect anomalies symptomatic of abusive software~\cite{RAPPOR, UnwantedSoftware,RapporBlogPost}.

Unfortunately,
in its current state,
the RAPPOR technology
can be of only limited applicability and utility.
This is because RAPPOR makes
two simplifying assumptions
that will certainly not always hold in practice:

\begin{trivlist}
\parshape 2 0cm \linewidth 2em \dimexpr\linewidth-2em\relax

\item
\textbf{Assumption~1: \emph{Aggregators only need to learn the distribution of a single variable, in isolation.}}
In practice, aggregators may want to study the association between multiple variables
because attributes are often more meaningful in association with other attributes.
For example,
in RAPPOR's application domain in the Chrome Web browser,
an innocent-looking homepage or search-provider URL
may become highly suspect if its use is
strongly correlated with installation of software that is known to be malicious.

\vspace*{1ex}

\item
\textbf{Assumption~2: \emph{Aggregators know the data dictionary of possible string values in advance.}}
There are many scenarios in which both the frequencies of client-side strings and the strings themselves may be unknown.
For instance, when collecting reports on installed software, it is unlikely that the names or hash values of all software will be known ahead of
time, especially in the face of polymorphic software.
Similarly, when studying user-generated data---manually-entered hashtags,
for instance---the dictionary of possible strings cannot be known \emph{a priori}.

\end{trivlist}

Lifting these two simplifying assumptions
is a significant challenge,
which requires reasoning about ``unknown unknowns.''
The first assumption
can only be removed by
estimating the unknown joint distributions of two or more unknown variables
that are observed only via differentially-private RAPPOR responses.
Removing the second assumption
requires learning a data dictionary of unknown client-side strings
whose frequency distribution is also unknown.
This process must additionally satisfy strong privacy guarantees
that preclude the use of encryption or special encodings
that could link individuals to strings.
Furthermore, neither of these challenges
admits a solution that is simultaneously feasible and straightforward.
The naive approach of trying all possibilities incurs exponential blowup over the
infinite domain of unknown strings, and is not even well-defined
with regards to estimating joint distributions.


This paper provides methods for addressing these two challenges,
thereby substantially improving upon
the recently-introduced RAPPOR statistical crowdsourcing technology.

First, regarding multivariate analysis,
we present a collection of statistical tools for studying the association between multiple random variables reported through RAPPOR.
This toolbox includes an expectation-maximization-based algorithm for inferring joint distributions of multiple variables from a collection of
RAPPOR reports.
It also includes tools for computing the variance of the distribution estimates, as well as testing for
independence between variables of interest.

Second,
regarding unknown data dictionaries,
we introduce a novel algorithm for estimating a distribution of strings without knowing the set of possible values beforehand.
This algorithm asks each reportee to send a noisy representation of multiple substrings from her string.
Using our previously-developed techniques for association analysis, we build
joint distributions of all possible substrings.
This allows the aggregator to learn the data dictionary for all frequent values underlying the reports.

If differential privacy is to gain traction outside the research community, we believe it is critical to tackle the practical challenges
that currently limit its immediate usefulness; in this work, we address two such challenges.
We demonstrate the practical efficacy of both contributions through simulation and real-world examples.
For these experiments we have created implementations of our analysis
that we are making publicly available (deferred for blind review).
While motivated by the recently-introduced RAPPOR mechanism, and presented in that context,
our contributions are not unique to the RAPPOR encoding and decoding algorithms.
Our methods can be easily extended to any locally differentially-private system
that is attempting to learn a distribution of discrete, string-valued random variables.

%% file: background.tex
\section{Background}
\label{sec:background}
A common method for collecting population-level
statistics without access to individual-level data points is based on \emph{randomized response}~\citep{warner}.
Randomized response is an obfuscation technique that satisfies a privacy guarantee known as \emph{local differential privacy}~\cite{localdiffpriv}.
We begin by briefly introducing local differential privacy and explaining how the RAPPOR system uses randomized response to
satisfy this condition.

Formally, a randomized algorithm $A$ (in this case, RAPPOR) satisfies $\epsilon$-differential privacy~\citep{dwork2006differential}
if for all pairs of client's values $x_1$ and $x_2$ and for all $R\subseteq Range(A)$,
$$
P(A(x_1) \in R) \le e^\epsilon P(A(x_2) \in R).
$$
Intuitively, this says that no matter what string user Alice is storing, the aggregator's knowledge about Alice's ground
truth does not change too much based on the information she sends. We would like to emphasize that
differential privacy is a property of an encoding algorithm, so these guarantees hold regardless of the underlying data
distribution.

RAPPOR is a privacy-preserving data-collection mechanism
that makes use of randomization to guarantee local differential privacy
for every individual's reports.
Despite satisfying such a strong privacy definition, RAPPOR enables the aggregator to accurately estimate a distribution over
a discrete dictionary (e.g., a set of strings).

The basic concept of \emph{randomized response} is best explained with an example.
Suppose the Census Bureau wants to know how many communists live in the United States without learning \emph{who} is a communist.
The administrator asks each participant to answer the question, ``Are you a communist?'' in the following manner:
Flip an unbiased coin. If it comes up heads, answer truthfully. Otherwise, answer `yes' with probability 0.5 and `no' with probability 0.5.
In the end, the Census Bureau cannot tell which people are communists, but it can estimate the true fraction of communists
with high confidence. Randomized response refers to this addition of carefully-designed noise to discrete random values
in order to mask individual data points while enabling the computation of aggregate statistics.

RAPPOR performs two rounds of randomized response to mask the inputs of users and enable the collection of user data over time.
Suppose Alice starts with the string $X$ (e.g., $X=\text{``rabbit'}$'). The sequence of events in the encoder is as follows:
\begin{enumerate}
\item Hash the string $X$ twice ($h$ times in general) into a fixed-length Bloom filter, $B$.
\item Pass each bit in the Bloom filter $B_i$ through a randomized response (giving $B_i'$) as follows:
$$
B'_i = \begin{cases}
1, & \text{with probability $\frac{1}{2}f$} \\
0, & \text{with probability $\frac{1}{2}f$} \\
B_i, & \text{with probability $1 - f$}
\end{cases}
$$
where $f$ is a user-tunable parameter controlling the level of privacy guarantees.
We refer to this noisy Bloom filter $B'$ as the \emph{permanent randomized response} (PRR) for the value $X$,
because this same $B'$ is to be used for both the current and all future responses about the value $X$.
\item Each time the aggregator requests a report, pass each bit $B_i'$ in the PRR through \emph{another} round of randomized response (giving $X_i'$),
as follows:
$$
P(X'_i = 1) = \begin{cases}
q, & \text{if $B'_i = 1$}. \\
p, & \text{if $B'_i = 0$}.
\end{cases}
$$
We refer to this array of bits $X'$ as an \emph{instantaneous randomized response} (IRR), and the aggregator only ever sees such bit vectors $X'$ for any value $X$ that is reported. The smaller the difference between $q$ and $p$ (user-tunable), the greater privacy guarantee is provided.
\end{enumerate}
This process is visualized in Figure \ref{fig:rappor}.

The RAPPOR encoding scheme satisfies two different $\epsilon$-differential
privacy guarantees: one against a one-shot adversary who sees only a single IRR, and one against a stronger adversary who sees infinitely
many IRRs over time. The latter adversary is able to reconstruct $B'$ with arbitrary precision after seeing enough reports, which motivates
the need for a PRR, but is unable to infer $B$ from a single copy of $B'$.
In principle, users could always report $B'$ at every data collection, but this would create a unique tracking identifier. 

If the set of possible strings is small and known prior to collection (e.g., country, gender, etc.), a simplified version of the algorithm,
called Basic RAPPOR, is more appropriate. The single difference is that in step (1), Basic RAPPOR does not make use of Bloom filters,
but deterministically assigns each string to its own bit ($h = 1$). In this case, the size of $B$ is determined by the cardinality of the set
being collected. This also significantly simplifies the inference process to estimate string frequencies by the aggregator.

Despite strong report-level differential privacy guarantees, RAPPOR can approximate the marginal distribution of the measured variable(s) with
high precision.
One high-utility decoding scheme is described in~\cite{RAPPOR}, 
but the details of marginal decoding are not critical to understanding of our present work.

%% file: regression.tex
\section{Estimating Joint Distributions}
Learning the distribution of a single variable is sometimes enough.
For example, if the aggregator's goal is to learn the 100 most popular URLs visited by clients, then a straightforward
application of the RAPPOR algorithm described in Section~\ref{sec:background} suffices.

More often, however, aggregators may be interested in learning the associations and correlations between \emph{multiple} variables,
all collected in a privacy-preserving manner.
For example, suppose we would like to understand the relationship between installed software and annoying advertisements, e.g., to detect the presence of so-called \emph{adware}.
To do so, we might study the association between displayed advertisements and recently-installed software applications or extensions.
If both of these variables are measured using the RAPPOR mechanism, the current literature does not describe how to estimate
their \emph{joint} distribution, although methods exist for estimating marginal frequencies of both variables individually.

In this section, we describe a general approach to estimating the joint distribution of two or more RAPPOR-collected
variables. Inference is performed using the expectation-maximization (EM) algorithm~\citep{em}, which produces unbiased
estimates of joint probabilities.
These joint estimation techniques will play a key role in Section \ref{sec:dictionary}, where we estimate data distributions over unknown dictionaries.

\subsection{Estimating Joint Distributions with the EM Algorithm}
The EM algorithm is a common way to obtain maximum likelihood estimates (MLEs) in the presence
of missing or incomplete data. It is particularly suited to RAPPOR applications where true values are
not observed (missing) and only their noisy representations are being collected.

For the sake of clarity, we will focus on estimating the joint distribution of two random variables $X$ and $Y$, both collected using Basic RAPPOR
introduced in Section~\ref{sec:background}. Extending this estimation to general RAPPOR requires careful consideration of unknown categories and will be discussed
in the next section.
Let $X' = \text{RAPPOR}(X)$ and $Y' = \text{RAPPOR}(Y)$ be the noisy representations of $X$ and $Y$ created by RAPPOR.
Suppose that $N$ pairs of $X'$ and $Y'$ are collected from $N$ distinct (independent) clients.

For brevity, let $X_i$ and $Y_j$ denote the events that $X=x_i$ and $Y=y_j$, respectively. The conditional probability of true values $X$ and $Y$, given the observed noisy representations
$X'$ and $Y'$, is just a consequence of Bayes' theorem:
$$
P(X = x_i, Y = y_j | X', Y') = \frac{p_{ij}P(X', Y' | X_i, Y_j)}{\sum\limits_{k = 1}^{m}\sum\limits_{\ell= 1}^{n} p_{k\ell}P(X', Y' | X_k, Y_{\ell})}.
$$
where $m$ and $n$ are the number of categories in $X$ and $Y$, respectively.
Here, $p_{ij}$ is the true joint distribution of $X$ and $Y$;
this is the quantity we wish to estimate for each
combination of categories $i$ and $j$. $P(X', Y' | X, Y)$ is the joint probability of
observing the two noisy outcomes given both true values. Because $X'$ and $Y'$ are conditionally independent given
both $X$ and $Y$, $P(X', Y' | X, Y) = P(X' | X)P(Y' | Y)$. Since the noise added through RAPPOR is predictable and mechanical,
it is easy to precisely describe these probabilities. Without loss of generality, assume that $X = x_1$. In Basic RAPPOR, $x_1$'s Bloom Filter representation has a one in the first position and zeros elsewhere, so we have
\begin{eqnarray*}
P(X' | X = x_1) &=& q^{x'_1}(1 - q)^{1 - x'_1}  \times p^{x'_2}(1 - p)^{1 - x'_2} \\
 	                && \times \ldots \times p^{x'_i}(1 - p)^{1 - x'_i}  \\
 		         && \times \ldots \times  p^{x'_m}(1 - p)^{1 - x'_m}.
\end{eqnarray*}

The EM algorithm proceeds as follows:
\begin{enumerate}
\item {\bf Initialize} $\hat{p}^0_{ij} = \frac{1}{nm}, 1 \le i \le m, 1 \le j \le n$ (uniform distribution).
\item {\bf Update} $\hat{p}^{t}_{ij}$ with
\begin{eqnarray*}
\hat{p}^{t+1}_{ij} & = & P(X = x_i, Y = y_j) \\
         & = & \frac{1}{N}\sum_{k = 1}^N P(X = x_i, Y = y_j | X'_k, Y'_k),
\end{eqnarray*}
where $P(X = x_i, Y = y_j | X'_k, Y'_k)$ is computed using the current estimates $\hat{p}^{t}_{ij}$.
\item {\bf Repeat} step 2 until convergence, i.e. $\max_{ij} | \hat{p}^{t+1}_{ij} - \hat{p}^t_{ij}| < \delta^*$ for some small positive value.
\end{enumerate}

This algorithm converges to the maximum likelihood estimates of $p_{ij}$, which are asymptotically unbiased.

\subsection{Handling the ``Other'' category}
In the EM initialization step, we assume that we know all $n$ categories
of $X$ and all $m$ categories of $Y$. In practice, the aggregator is unlikely to know all the
relevant categories, and must make choices about which
categories to include.
Operationally, the aggregator would perform marginal analyses on both $X'$ and $Y'$ separately, estimate
the most frequent categories, and use them in the joint analysis. The remaining undiscovered categories, which we refer to as 
``Other'', cannot be simply omitted from the joint analysis because doing so leads to badly biased distribution estimates. In this section, we discuss how to handle this problem.

Suppose one ran the marginal decoding analysis separately on $X'$ and $Y'$, thereby detecting $m$  and $n$ top categories, respectively, along with their corresponding marginal frequencies. Note that $m$ and $n$ now represent the \emph{detected} numbers of categories instead of the true numbers of categories.
The ``Other'' categories for $X$ and $Y$ 
may constitute a significant amount of probability mass (computed as $1 - \sum_{i = 1}^{m}\hat{p}_i$ and $1 - \sum_{j = 1}^{n}\hat{p}_j$, respectively)
which must be taken into account when estimating the joint distribution.

The difficulty of modeling the ``Other'' category comes from the apparent problem of estimating
\begin{equation}
P(X' = x' | X = ``Other\text{''}),
\label{eq:other}
\end{equation}
i.e. the probability of observing a report $x'$ given that it was generated by any category other than the top $m$ categories of $X$.
However, if we could estimate this probability we could simply use the EM algorithm
in its current form to estimate the joint distribution---an $(m + 1) \times (n + 1)$ contingency table in which the last row and the last column are the ``Other'' categories for each variable.

We use knowledge of the top $m$ categories and their frequencies to estimate the probability in (\ref{eq:other}).
Let $c_{s}^m$ be the expected number of times that \emph{reported} bit $s$ was set by one of the top $m$ categories in $X$. It is equal to
$$
c_{s}^m =  \left(\left(1 - \frac{f}{2}\right)q + \frac{fp}{2}\right)T_{s} + \left(\left(1 - \frac{f}{2}\right) p + \frac{fq}{2}\right)(N - T_{s}),
$$
where $N$ is the number of reports collected and
$$
T_{s} = N\sum_{i = 1}^m p_iI(B_{s}(x_i) = 1)
$$
represents the expected number of times the $s$th bit in $N$ Bloom filters was set by a string
from one of the top $m$ categories.
Here, $I$ is the indicator function returning 1 or 0 depending if the condition is true or not,
$B(x_i)$ is the Bloom filter generated by string $x_i$ and $p_i$ is the
true frequency of string $x_i$.

Given the above, the estimated proportion of times each bit was set by a string from the ``Other'' category is then
$$
\hat{p}^o_{s} = \frac{c_{s} - \hat{c}_{s}^m(\hat{p}_i)}{N(1 - \sum_{i = 1}^m \hat{p}_i)},
$$
where $c_s$ is the observed number of times bit $s$ was set in all $N$ reports.

Then, the conditional probability of observing any report $X'$ given that the true value was ``Other'' is given by
$$
P(X' = x' | X = ``Other\text{''}) = \prod_{s = 1}^k \left(\hat{p}_{s}^o\right)^{x'_s} \left(1 - \hat{p}_{s}^o\right)^{1 - x'_s}.
$$

As stated earlier, we can use this estimate to run the EM algorithm with ``Other" categories, thereby obtaining unbiased estimates of the joint distribution.

\subsection{Estimating the Variance-Covariance matrix}
It is a well-known fact in statistics that the asymptotic distribution of the maximum likelihood estimates $(\hat{p}_{11}, \ldots, \hat{p}_{mn})$ is
$$
N\left((p_{11}, \ldots, p_{mn}), I^{-1}\right),
$$
where $N(\mu, \Sigma)$ stands for a Gaussian distribution with mean $\mu$ and variance-covariance matrix $\Sigma$ and $I$ is the information
matrix defined below.
(See~\cite{mle}.) 
A good estimate of
$\Sigma$ is critical, as it allows us to assess how certain we are about our estimates of $p_{ij}$'s.
It permits an aggregator to construct 95\% confidence intervals, rigorously test if any of the proportions are different from 0, or
perform an overall test for the association between $X$ and $Y$.

In this case, the asymptotic variance-covariance matrix is given by the inverse of incomplete-data observed information matrix $I_{obs}$.
To obtain an estimate of the information matrix, we would evaluate the second derivative of the observed-data log-likelihood function
at our MLE estimates $\hat{p}_{ij}$'s.

The log-likelihood function is the log of the probability of observing all $N$ reports, treated as a function of the unknown
parameter vector $(p_{11}, \ldots, p_{mn})$: 
$$
l(p_{11}, \ldots, p_{mn}) = \sum_{k = 1}^N\log\left(\sum_{i = 1}^{m}\sum_{j = 1}^n p_{ij}P(X'_k, Y'_k | X_i, Y_j )\right).
$$
The first derivative with respect to $p_{ij}$ is given by
$$
l'_{p_{ij}} = \sum_{k = 1}^N\frac{P(X'_k, Y'_k | X = x_i, Y = y_j)}{\sum_{o = 1}^{m}\sum_{\ell = 1}^n p_{ij}P(X'_k, Y'_k | X = x_{o}, Y = y_{\ell})}.
$$

The second derivative, also known as the observed information matrix (size $mn \times mn$), is given by
$$
l''_{p_{ij}, p_{st}} = \sum_{k = 1}^N\frac{P(X'_k, Y'_k |x_i, y_j) \cdot P(X'_k, Y'_k | x_s, y_t)}{\left(\sum_{o = 1}^{m}\sum_{\ell = 1}^n p_{ol}P(X'_k, Y'_k | x_o, y_\ell)\right)^2}.
$$
Inverting this matrix and evaluating at the current MLE estimates $\hat{p}_{11}, \ldots, \hat{p}_{mn}$ results in an estimate of the 
variance-covariance matrix $\hat{\Sigma}$. The $mn$ diagonal elements of $\hat{\Sigma}$ contain the variance estimates for each $\hat{p}_{ij}$ and
can be directly used to assess how certain we are about them.

\subsection{Testing for Association}
When dealing with two or more categorical variables, one of the first questions generally asked is whether they are independent of each other.
For two variables to be independent their joint distribution must be equal to the product of their marginals, i.e.
$$
P(X, Y) = P(X | Y)P(Y) = P(Y | X)P(X) = P(X)P(Y).
$$
In practice, this means that knowing the value of $X$ provides no valuable information when predicting $Y$.
In this section, we explain why the most common statistical test of independence, the $\chi^2$ test~\cite{agresti}, is not appropriate for
RAPPOR-collected variables and propose an alternative test statistic.

The $\chi^2$ test is one of the most widely used statistical techniques for testing the independence
of two or more categorical variables. It proceeds by comparing the observed cell counts to what is expected under the
independence assumption. The formal test statistic is given by
$$
\chi^2 = \sum_{i=1}^{mn} \frac{(O_i - E_i)^2 }{E_i},
$$
where $E_i$ is expected number of cell counts under the independence assumption and $O_i$ is the observed number of cell counts. 
This test statistic
has a known distribution under the assumption that $X$ and $Y$ are independent, and it is a $\chi^2$ distribution with $(m-1)(n-1)$ degrees of freedom.

Unfortunately, we cannot use the $\chi^2$ test statistic because we do not observe exact cell counts $O_i$ of the co-occurrence of our random variables $X$ and $Y$.
Instead, we have mean estimates and the corresponding variance-covariance matrix obtained through the EM algorithm.

Weighted quadratic forms of multivariate normal inputs are well-behaved with tractable distribution properties. 
Let
$$
T= (\hat{p} - \hat{\mu})^\mathsf{T} \hat{\Sigma}^{-1} (\hat{p} - \hat{\mu}),
$$
where $\hat{p}$ is a vector of $\hat{p}_{ij}$'s, $\hat{\mu}$ is a vector of products of marginals (i.e. the expected joint distribution if the variables are independent) and $\hat{\Sigma}$ is the estimated variance-covariance matrix. Here, $\mathsf{T}$ indicates the transpose operation.
Under the null hypothesis of no association, this test statistic $T$ has a $\chi^2$ distribution with $(m-1)(n-1)$ degrees of freedom, similarly
to the $\chi^2$ test.

In summary, to perform a formal statistical test for independence between $X$ and $Y$, one would use the EM algorithm to estimate
the joint distribution along with the variance-covariance matrix. Then, one would compute the $T$ test statistic and compare it to the corresponding
critical quantile $q_{1-\alpha}$ from the $\chi^2_{(m-1)(n-1)}$. We would conclude that $X$ and $Y$ are not independent if $T > q_{1-\alpha}$ and
state that there is no evidence for non-independence otherwise.
We demonstrate numerically that this proposed test statistic has the expected behavior in Appendix \ref{app:independence}.

\begin{figure*}[htbp]
\begin{center}
\includegraphics[scale=.6]{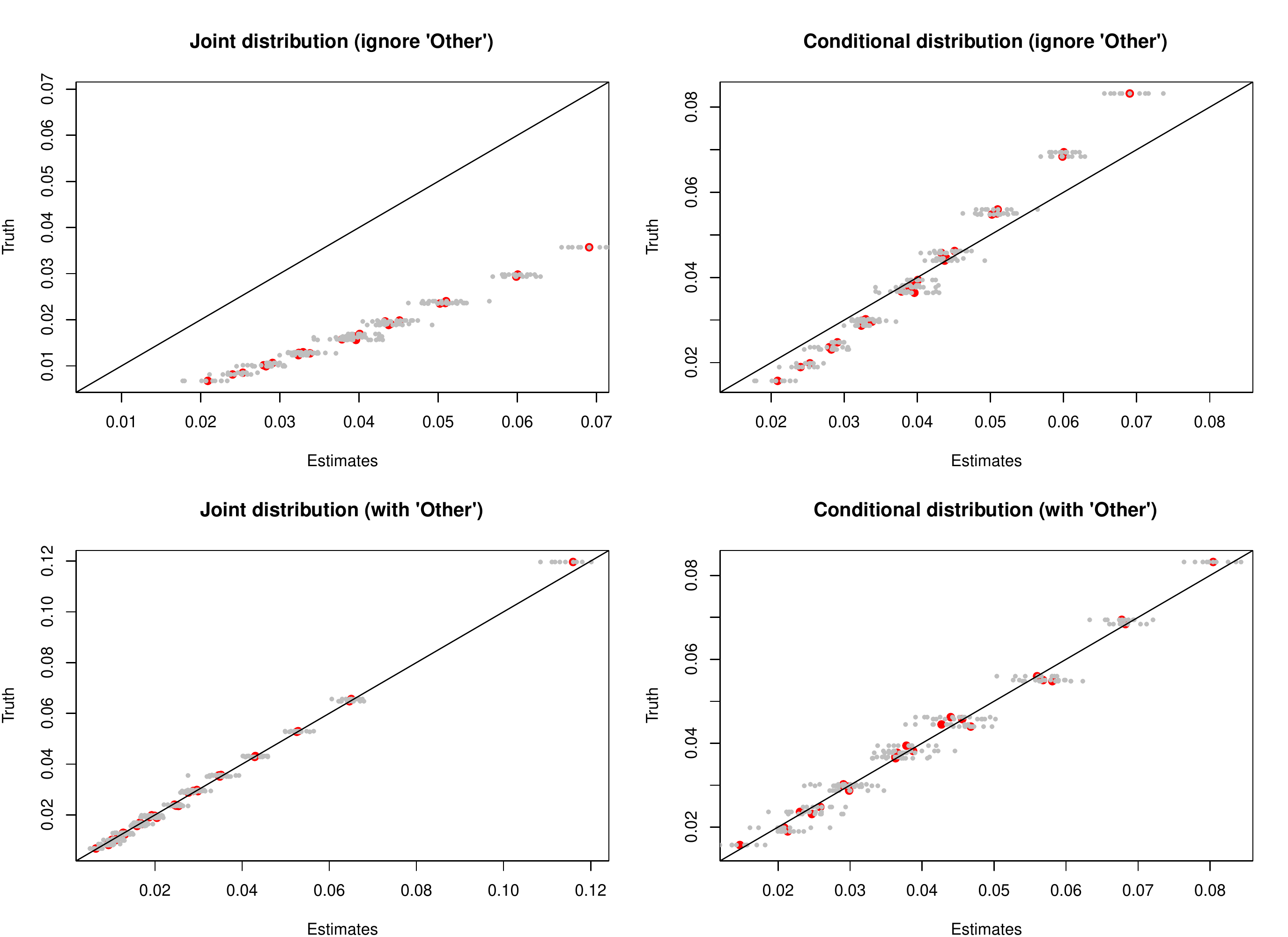}
\caption{Sample size 100,000. True vs. estimated joint frequencies. Red dots show average estimates over 10 Monte Carlo runs.
Grey dots show individual estimates. Two panels at the top show how ignoring the ``Other'' category leads to biased estimates for
both joint and conditional distributions. For the conditional distribution, there's a regression to the mean effect where high values
are underestimated and lower values are overestimated.
Accounting for the ``Other'' category fixes the problem and the estimates
are close to the truth.}
\label{fig:other}
\end{center}
\end{figure*}

\begin{table}[!t]
\begin{center}
\caption{True joint distribution of $X$ (in rows) and $Y$ (in columns).}
\label{tab:rapportruth}
\begin{tabular}{rrrrrrr}
  \hline
 & 1 & 2 & 3 & 4 & 5 & Other \\
  \hline
1 & 3.567 & 2.937 & 2.468 & 1.952 & 1.639 & 6.436 \\
   2 & 2.984 & 2.432 & 1.967 & 1.581 & 1.289 & 5.362 \\
   3 & 2.473 & 1.991 & 1.609 & 1.223 & 1.025 & 4.343 \\
   4 & 1.881 & 1.569 & 1.293 & 1.069 & 0.874 & 3.499 \\
   5 & 1.625 & 1.292 & 1.080 & 0.892 & 0.662 & 2.836 \\
   Other & 6.380 & 5.292 & 4.311 & 3.495 & 2.809 & 11.863 \\
   \hline
\end{tabular}
\end{center}
\end{table}

 \begin{table}[!t]
 \centering
\caption{Estimated joint distribution of $X$ (in rows) and $Y$ (in columns). }
\label{tab:rapporest}
 \begin{tabular}{rrrrrrr}
   \hline
  & 1 & 2 & 3 & 4 & 5 & Other \\
   \hline
 1 & 3.306 & 2.952 & 2.413 & 2.018 & 1.806 & 6.691 \\
   2 & 3.045 & 2.292 & 2.043 & 1.588 & 1.286 & 5.302 \\
   3 & 2.336 & 2.173 & 1.587 & 1.115 & 0.916 & 4.450 \\
   4 & 1.902 & 1.506 & 1.354 & 1.087 & 0.887 & 3.510 \\
   5 & 1.763 & 1.233 & 1.188 & 0.873 & 0.615 & 2.801 \\
    Other & 6.531 & 5.338 & 4.245 & 3.513 & 2.916 & 11.419 \\
    \hline
 \end{tabular}
 \end{table}
 
\subsection{Simulation Results}
To illustrate our multivariable analysis of differentially private data,
we generated synthetic RAPPOR reports for variables $X$ and $Y$, each with 100 unique categories.
The marginal distributions of $X$ and $Y$ were
discretized Zipfian distributions, and their (truncated) joint distribution is given in Table \ref{tab:rapportruth}.

With 100,000 reports we were able to estimate the frequencies of 15 top categories for each $X$ and $Y$, on average. For the purpose of performing the association analysis, we
selected the top five categories from each variable as indicated by the estimated marginal distribution. 
First we ignored the ``Other'' categories completely and assumed that $X$ and $Y$ had 5 unique values each.
10 Monte Carlo replications were performed; the first two panels of Figure \ref{fig:other} plot the estimated cell frequency against the true
cell frequency for each of the ten trials and 25 cells.
As expected, the estimated 25 proportions are poor estimates for both the true joint frequencies and the conditional frequencies
$P(X = x, Y = y | X \in \text{top-5}, Y \in \text{top-5})$. In fact, for the conditional probabilities, there is a regression to the mean effect
where high values are under-estimated and low values are overestimated.

The bottom two panels of Figure \ref{fig:other} show estimates when we account for the ``Other'' categories of both $X$ and $Y$.
The estimated joint distribution is now a $6 \times 6$ table,  and the procedure produces unbiased estimates for the true joint frequencies (Table \ref{tab:rapporest}).
Accordingly,
it also produces 25 unbiased estimates for conditional frequencies.

\subsection{Real-World Example: Google Play Store Apps}
To demonstrate these techniques in a real example, we downloaded the public
metadata for 200,000 mobile-phone apps from the Google Play Store.
For each of these apps, 
we obtained the app category (30 categories) and whether it is offered for free or
not. 
This information can be summarized in a $30 \times 2$ contingency table.
Applying a $\chi^2$ independence test to this contingency table would test whether
different categories are statistically more likely to feature free apps. 
In this section, we use RAPPOR and our joint decoding approach to learn this distribution
without direct access to the underlying data points.
The dataset in this example is not particularly sensitive; however, we were unable to find public
datasets of sensitive, multivariate, categorical data, precisely due to 
the associated privacy concerns.

For each sampled app, we generated a simulated Basic RAPPOR report for both variables: app category and payment model.
We used 30-bit reports for the category variables, and 1-bit reports for the Boolean payment model.

\begin{figure*}[htbp]
\begin{center}
\includegraphics[scale = .52]{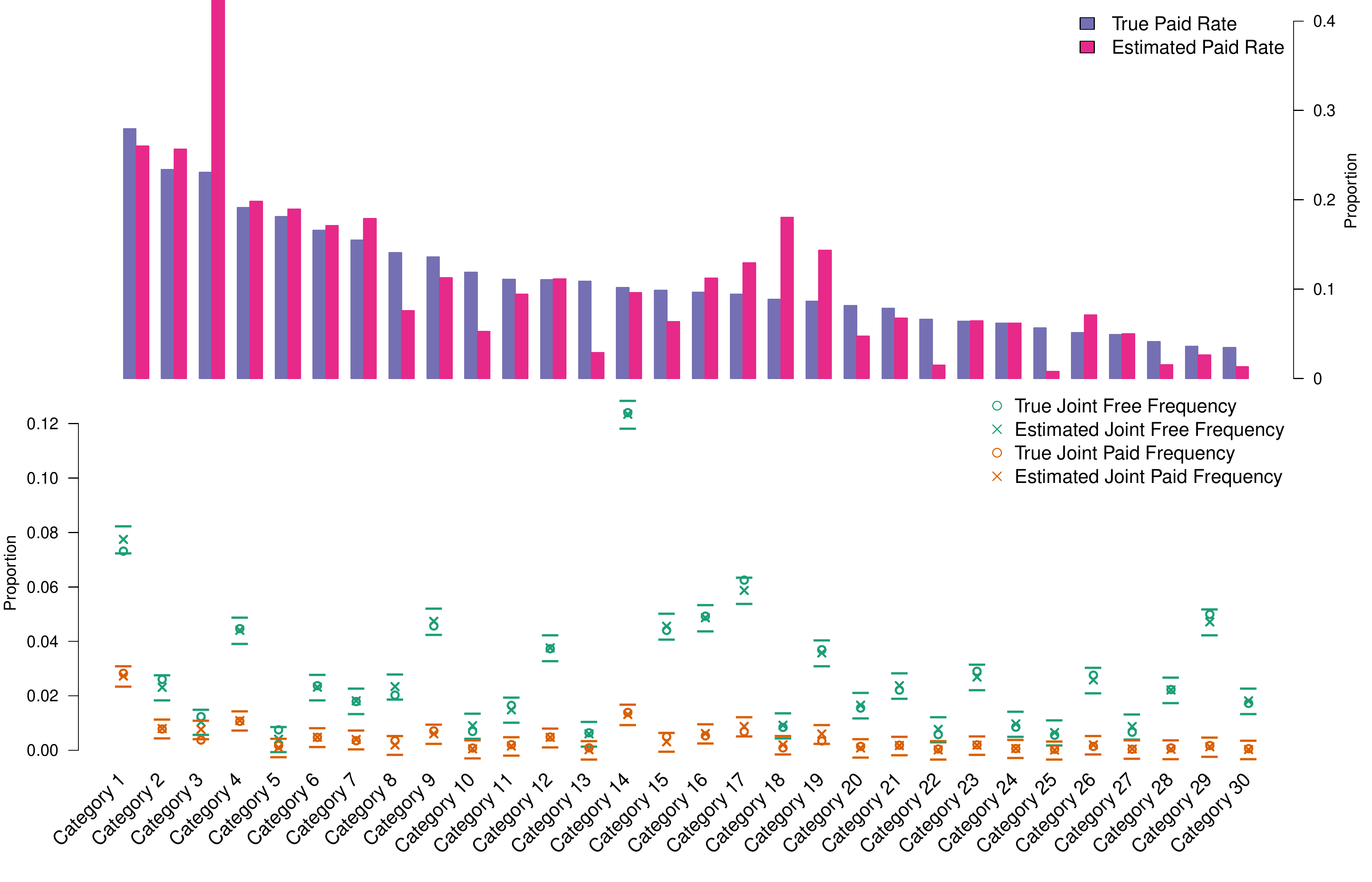}
\caption{Estimating the joint distribution of the categories of software items, and whether they are free or for purchase. Categories are ordered by the paid fraction, shown in the top panel.
The bottom panel plots 60 true and estimated \emph{joint} frequencies along with 95\% confidence intervals shown as horizontal bars. Both true and estimated frequencies (free + paid) add up to 1.}
\label{fig:ps}
\end{center}
\end{figure*}

We then performed a joint distribution analysis by estimating the $30 \times 2$ contingency table---i.e., the frequency of each combination
of item category and payment model. Results are shown in the second panel of Figure \ref{fig:ps}. The green points show both true and estimated
frequencies of free items for each category, while the brown points show the paid ones.
Note that these are the 60 cell frequencies from the true and estimated contingency tables, not proportions of free or paid apps for each category.
95\% confidence intervals are shown as horizontal bars for both
sets of estimates and have proper coverage in all cases.

The top panel of Figure \ref{fig:ps} shows the true and estimated paid rate for each category, computed as the proportion of paid apps for that category
divided by the overall proportion of a category. This ratio estimate is less stable than the joint frequencies but follows the true rates
closely for most app categories.

We perform a formal test for independence by computing the proposed $\chi^2$-test statistic $T = 107.093$, which has a p-value of
$6.9523e-11$. This is much smaller than $0.05$ and we would therefore conclude that there are, in fact, statistically significant differences
in paid rates between different app categories. This can be, of course, clearly seen from the top panel where categories are ordered in the descending
prevalence of paid software, with proportions ranging from 30\% to 4\%.

%% file: appendices.tex
\appendices

\section{Testing for Association: Validation}
\label{app:independence}
We wish to demonstrate the validity of our proposed statistic for testing the independence of two
or more variables collected with RAPPOR.

\begin{figure}[htbp]
\begin{center}
\includegraphics[width = 3.5in]{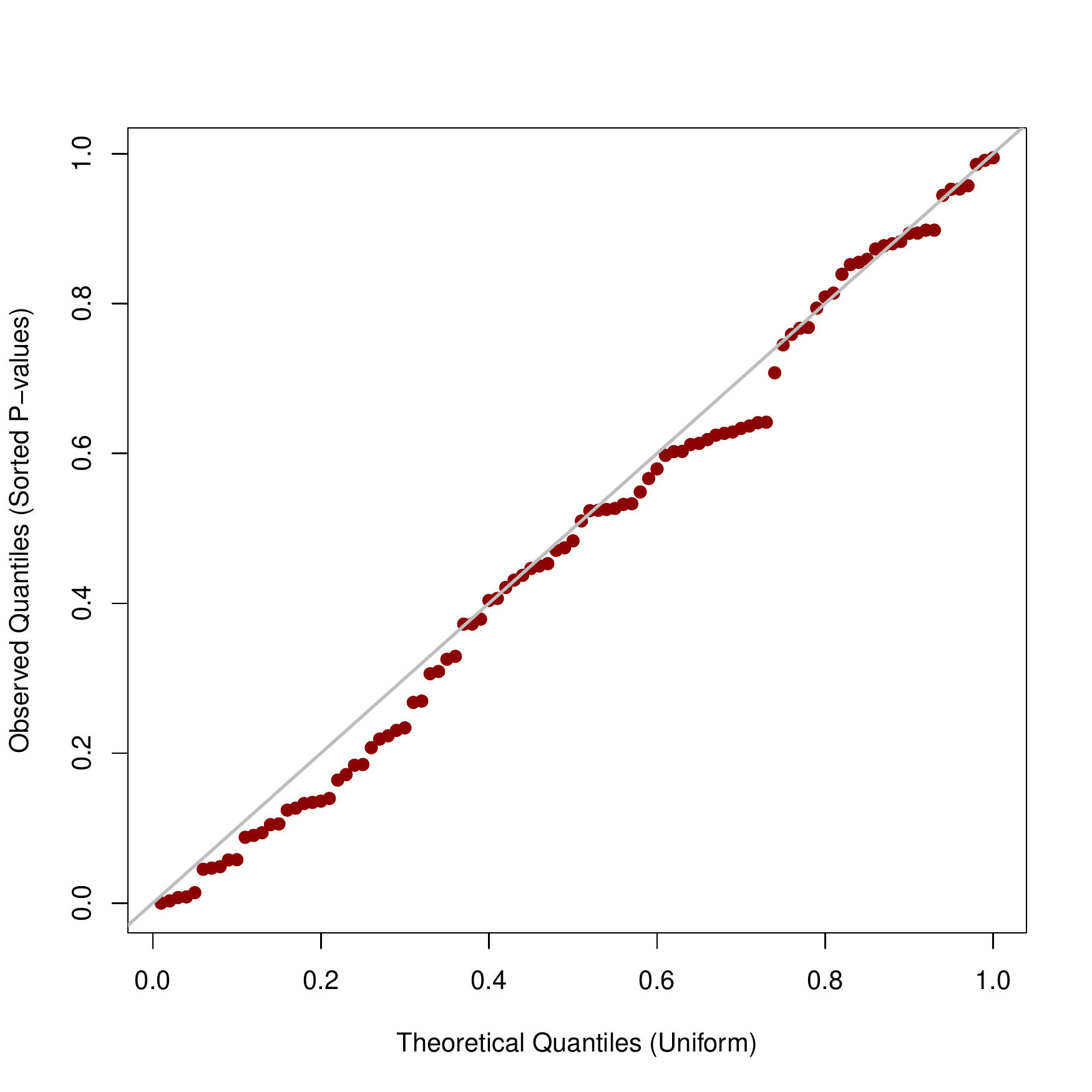}
\caption{Quartiles.}
\label{fig:quartiles}
\end{center}
\end{figure}

Consider two independent random variables, $X$ and $Y$.
If we test a null hypothesis on these variables at a confidence level of $\alpha=0.05$,
we would expect to falsely reject the null hypothesis 5 percent of the time.
More precisely, if the test statistic is continuous (as ours is), then the $p$-value is uniformly distributed between 0 and 1 if the null hypothesis is true.

Thus, in order to demonstrate that our proposed statistic can be used as a test of independence,
we generated a pair of distributions of independent random variables, $X$ and $Y$.
In each trial, we drew $K = 10,000$ data points $\{(x_1,y_1), \ldots, (x_{K},y_{K})\}$.
After encoding these data points with RAPPOR and then jointly decoding the reports, we obtain estimates $\hat p_{ij}$ and $\hat \Sigma$.
We use these estimates to compute our proposed statistic, $T$, for a single trial.
We did this for 100 trials.

Since the null hypothesis is true by construction, the $p$-values of our test should have a distribution that is uniform. Therefore, the expected quantiles of our constructed dataset should be uniformly spaced between 0 and 1.
Figure \ref{fig:quartiles} plots these expected quantiles against our observed quantiles from the dataset. Because the points are well-represented by a linear fit with slope 1 and intercept 0,
we conclude that our test statistic has the desired properties as a test of variable independence.

\section{Learning Unknown Dictionaries: \\ Different Distributions}
\label{app:distributions}
\begin{figure}[htbp]
\begin{center}
\includegraphics[width = 3.5in]{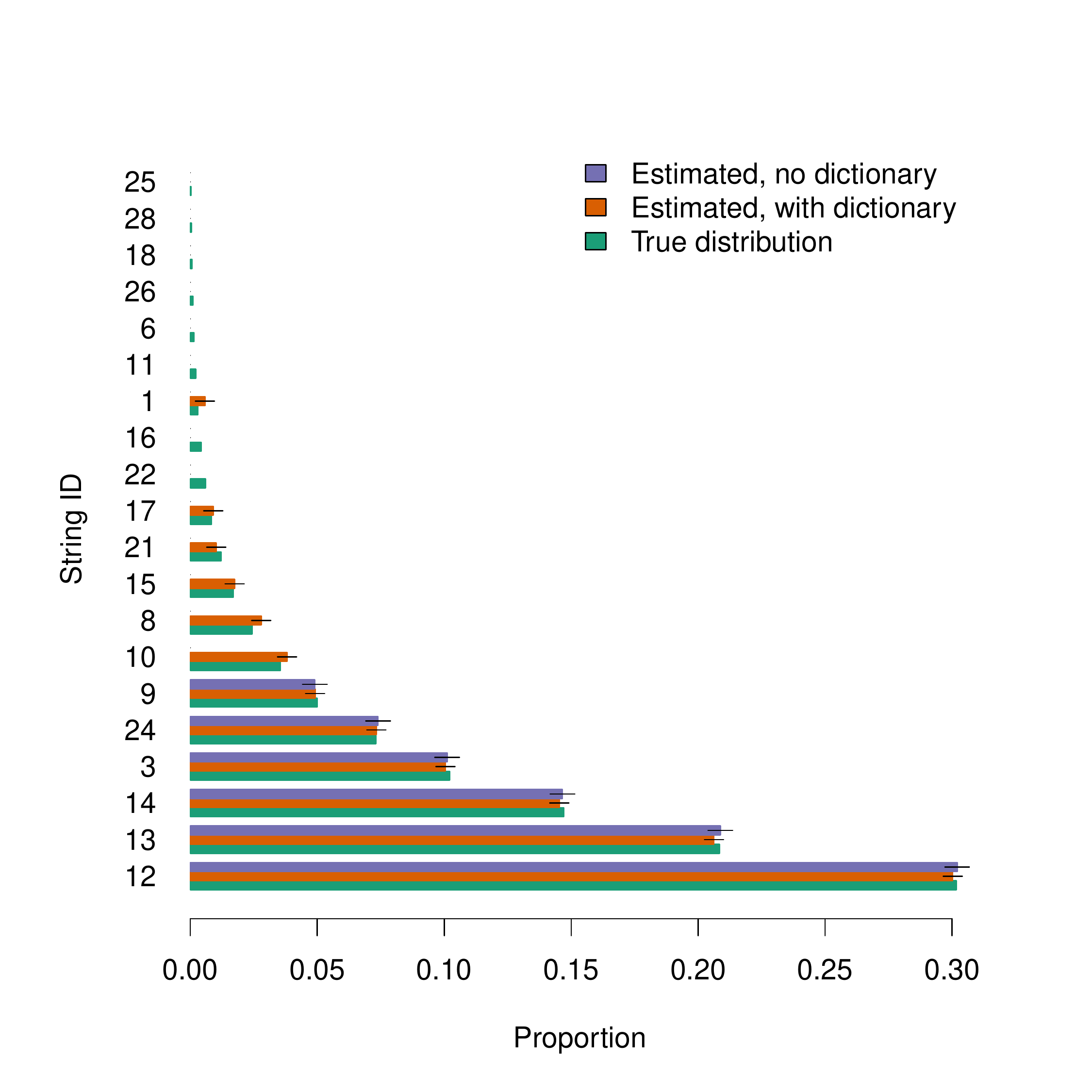}
\caption{Estimated distribution of hash strings without knowing the dictionary a priori.
The underlying distribution is geometric with $p=0.3$.}
\label{fig:geometric}
\end{center}
\end{figure}

Throughout this paper, our simulations were run over discrete approximations of
Zipfian distributions.
This class of distributions is common in practice, but for completeness, we also
include results from other distributions that may arise in practice.
All plots in this section were generated by running the decoding algorithm over
$N=100,000$ simulated reports.
Each report was generated from a string drawn from a specified distribution over
100 categories.
The privacy parameters used were $p=0.25$, $q=0.75$, and $f=0$,
and we used 128-bit Bloom filters.
Figure \ref{fig:geometric} shows the estimated distribution when the underlying
distribution is a truncated geometric with parameter $p=0.3$.
The geometric distribution is the discrete equivalent of an exponential
distribution.
In terms of estimate accuracy, there is little difference between this
distribution and the Zipfian distribution we used in simulation.

\begin{figure}[htbp]
\begin{center}
\includegraphics[width=3.5in]{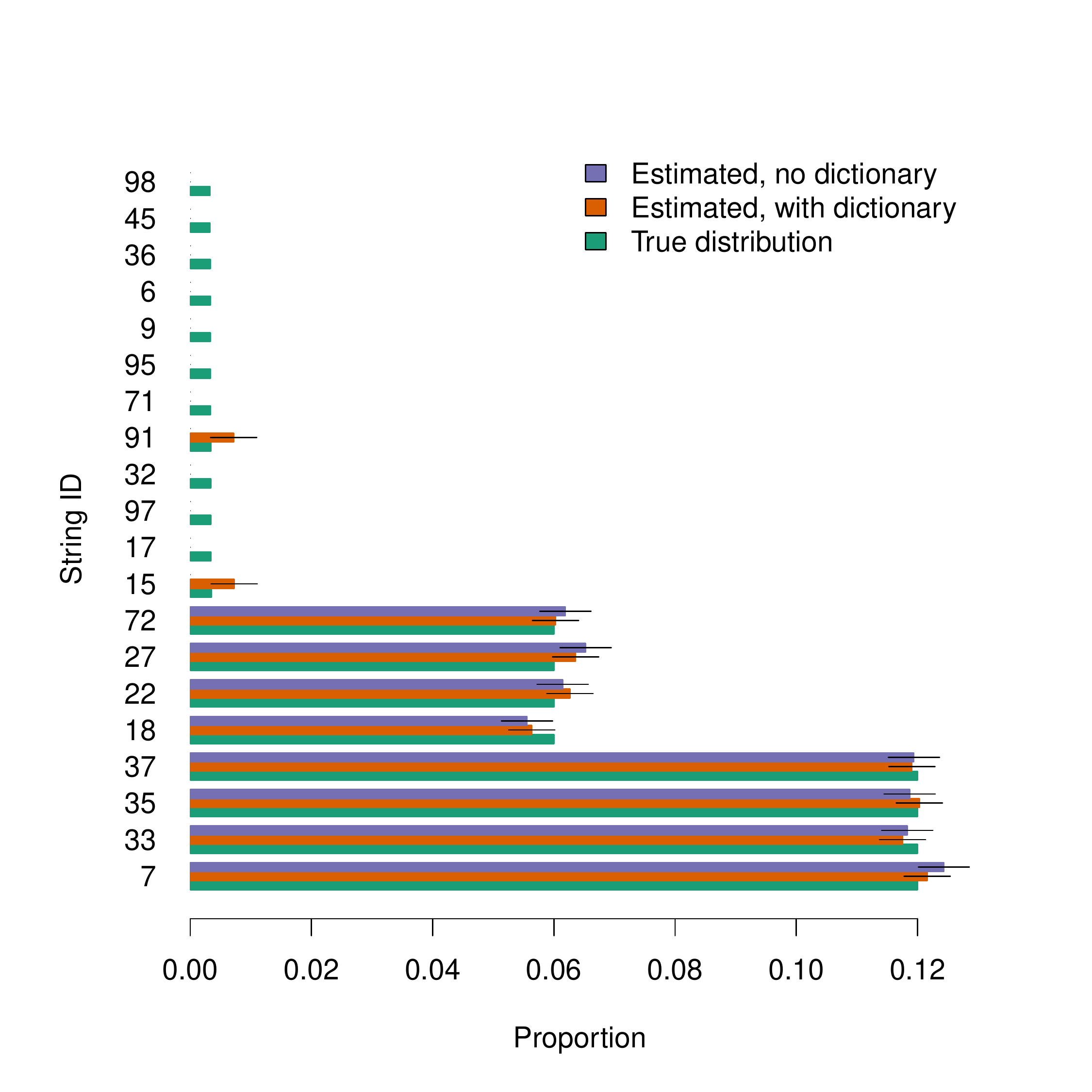}
\caption{Estimated distribution of hash strings without knowing the dictionary a priori.
The underlying distribution is a synthetic step function we constructed.}
\label{fig:step}
\end{center}
\end{figure}

Figure \ref{fig:step} shows the estimated distribution when the underlying
distribution is a synthetic stepwise function.
This function has four strings with probability mass 0.12, four strings with
probability mass 0.06, and the remaining 92 strings share the remaining mass
evenly.
In this distribution we can see that the estimates are less accurate than
in the geometric distribution.
This occurs because in the step distribution, a significant fraction of probability
mass lies below the noise threshold and gets accumulated in the ``Other'' category.
While our approach to handling the ``Other'' category leads to unbiased estimates,
it nonetheless relies on approximations from noisy data.
Therefore, the more significant the ``Other'' category, the less certain we are about
the bigram joint frequencies, leading to worse estimates overall.
The takeaway message is that the flatter the distribution (i.e. the greater the entropy), the
lower the likelihood of our algorithm accurately capturing the distribution.
Indeed, if we run this algorithm on a uniform distribution with each string having
probability 0.01, \emph{none} of the strings are found.

